\documentclass[a4paper,11pt]{article}
\usepackage{jcappub} % for details on the use of the package, please
\usepackage[T1]{fontenc} % if needed
\usepackage{subfigure}
\usepackage[normalem]{ulem}

\usepackage{enumitem,amssymb,amsmath}

\def \O{\Omega}

\def \f{\frac}

\def \a{\alpha}

\def \O{\Omega}

\def \p{\partial}
\def \d{\Delta}

\def \a{\alpha}

\def \th{\theta}

\def \ra{\rightarrow}

\title{Study of a Tilted Thin Accretion Disk around a Kerr-Taub-NUT black hole}

\author[a]{Gargi Sen,}
\author[b]{Chandrachur Chakraborty,}
\author[c]{Sudip Bhattacharyya}
\author[a]{Debaprasad Maity,}
\author[a]{Sayan Chakrabarti,}
\author[a]{Santabrata Das}
\affiliation[a]{Department of Physics, Indian Institute of Technology Guwahati,\\ Guwahati, 781039 India}
\affiliation[b]{Manipal Centre for Natural Sciences, Manipal Academy of Higher Education, \\ Alevoor Road, Manipal, 576104 India}
\affiliation[c]{Department of Astronomy and Astrophysics, Tata Institute
of Fundamental Research,\\ 1 Homi Bhabha Road, Colaba, Mumbai, 400005 India}
\emailAdd{g.sen@iitg.ac.in}
\emailAdd{chandrachur.c@manipal.edu}
\emailAdd{sudip@tifr.res.in}
\emailAdd{debu@iitg.ernet.in}
\emailAdd{sayan.chakrabarti@iitg.ac.in}
\emailAdd{sbdas@iitg.ac.in}

\abstract{The accreting collapsed object GRO J1655-40 could contain the gravitomagnetic monopole (GMM), and it was shown to be better described by the Kerr-Taub-NUT (KTN) spacetime instead of the Kerr spacetime. The warped accretion disk has also been observed for the same collapsed object. Motivated by these, we study a tilted thin inner accretion disk around a KTN black hole. Such a tilting could have a significant effect on the X-ray spectral and timing features via the Lense-Thirring effect. Taking into account the contribution from the inner accretion disk for the KTN black hole, here we calculate the radial profile of a tilt angle. Depending on the numerical values of the viscosity of the accreting material and Kerr parameter, GMM tends the angular momentum of the disk to align along the black hole's spin axis, or to make it more tilted. Our solution for the radial profile of the tilted disk around a KTN black hole could be useful to probe the strong gravity regime, and could also give indirect evidence for the existence of GMM in nature.}


\begin{document}

\maketitle
\flushbottom

\section{Introduction}\label{intro}

Orbital plane of a test particle around a central rotating object precesses due to the Lense-Thirring (LT) effect \cite{lt}. One cannot see such a similar effect in the Schwarzschild spacetime, which apparently indicates that the LT effect arises due to the rotation of the spacetime. If one goes deeper, one can discover that the `rotation' is not the fundamental entity responsible for the orbital plane precession. It turns out that the orbital plane precession can arise in any stationary spacetime \cite{cm} which violates the time reflection $(t \ra -t)$ symmetry but preserves the time translation $(t \ra t+k$, where $k$ is a constant) symmetry \cite{wald}. The presence of spin paramater or Kerr parameter ($a$) is responsible to make the Kerr spacetime stationary. In the absence of Kerr parameter, the Kerr spacetime reduces to the Schwarzschild spacetime which is geometrically static and spherically symmetric. If this {\it non-rotating} Schwarzschild spacetime contains the NUT (Newman-Unti-Tamburino) \cite{nut} parameter/charge, it is called as the Taub-NUT spacetime which is stationary but spherically symmetric \cite{mis}. As the Taub-NUT spacetime violates the time reflection symmetry, it is stationary. Furthermore, like in the Schwarzschild case, the Taub-NUT spacetime is spherically symmetric since the three linearly independent spacelike Killing vectors satisfy the commutation relation (see eq. (5) or eq. (18) of  \cite{mis} for details). In principle, the stationarity of the Taub-NUT spacetime allows us to observe the orbital plane precession \cite{Chakraborty-Bhattacharyya2018}. In this context, one should note here that the orbital plane precession even arises in the magnetized Schwarzschild spacetime  \cite{glp} (which is static and axisymmetric \cite{gmptn}), surpassing earlier folklore that only a stationary spacetime can generate such a precession. This magnetic field generated phenomenon is known as Gravitational Larmor precession. Thus, it is evident that the orbital plane does not only precess in the stationary and axisymmetric spacetime, but it can also precess if either of them does not hold. One should note here that the NUT charge $(n)$ is physically interpreted as `a linear source of pure angular momentum' \cite{bon, dow}, and it is also known as the `dual mass' or gravitomagnetic monopole (GMM) \cite{Chakraborty-Bhattacharyya2018}. $n$ is not like the intrinsic angular momentum of a rigid body. It, in fact, gives a `rotational sense' \cite{cm} in this spacetime, as it violates the time reflection symmetry. Now, if a Kerr spacetime contains the NUT parameter/GMM, it is called as the Kerr-Taub-NUT (KTN) spacetime. 
The immediate question that arises, how the orbital plane precession or so-called LT effect is influenced by the presence of an `intrinsic rotation' (i.e., $a$) and a `rotational sense' (i.e., $n$). Additionally, what would be the observational consequences of the presence of these two different types of angular momentum parameters?

When a black hole accretes matter from a distant star, an accretion disk is formed around the black hole. The accretion disk formed around a spinning black hole experiences the LT precession. Bardeen and Petterson suggested \cite{bp} that the LT effect could change the disk structure around the spinning black hole, if the accretion disk is not in the equatorial plane with respect to the black hole spin axis.
Since the LT precession frequency varies with $R^{-3}$ (where $R$ is the distance from the central object), the LT torque dominates over the viscous torque of the accreting material close to the black hole. On the other hand, the viscous torque dominates over the LT torque far away from the black hole. The tug-of-war between these two torques divides the accretion disk into three regions. The outer part of the disk remains tilted, the inner part aligns along the equatorial plane due to the strong LT effect and a twisted transition region forms between the above-mentioned two regions. The alignment of the inner accretion disk in the equatorial plane around a spinning black hole is known as the Bardeen-Petterson (BP) effect. The original formulation of Bardeen and Petterson was modified several times by others to remove some inconsistencies \cite{hbs, p71, p72, p73}. 

Papaloizou and Pringle \cite{Papaloizou-Pringle1983} first derived the equations for a viscous, tilted accretion disk under external forces, considering two kinematic viscosities, i.e., shear within the plane and perpendicular to the disk, denoted by $\nu_1$ and $\nu_2$, respectively. 
Later, Pringle \cite{Pringle1992} introduced a more general formalism of the warped disk equations based on the conservation of angular momentum. This was broadly
supported by the later detailed analysis of
Ogilvie \cite{Ogilvie-1999} who introduced a third viscosity ($\nu_3$) along with $\nu_1$ and $\nu_2$. 
The linear hydrodynamical analysis of \cite{Papaloizou-Pringle1983} was later expanded to the nonlinear regime by \cite{Ogilvie-1999}, leading to the evolution equations for any arbitrary warp amplitudes. This derivation confirmed the equations
obtained by Pringle \cite{Pringle1992} in the viscous regime \cite{Banerjee-etal2019}.
Later, Scheuer and Feiler \cite{sf} introduced the LT effect in the Pringle's equations and solved the warp disk equation. However, the contribution from the inner accretion disk was not taken into account. Chakraborty and Bhattacharya \cite{cb} first analytically solved the full warp disk equation to obtain the expression of tilt angle upto the first order in $a$ taking into account the inner disk contribution in the warp disk equation derived by Scheuer and Feiler. Banerjee et al. \cite{Banerjee-etal2019} solved the same warp disk equation and showed that the inner
disk may not be aligned at all for certain reasonable ranges of parameter values. It was timely because Ingram et al. \cite{in} discovered a tilted inner accretion disk around the black hole H1743-322 from the astrophysical observation. Note that the above-mentioned warped disk scenario has been probed assuming that the black hole is described by the Kerr geometry.

As previously mentioned, the LT effect or orbital plane precession, which is the fundamental cause of the tilted and warped disk,  not only appears in the Kerr spacetime but also in the KTN spacetime. Thus, one can study a tilted thin inner accretion disk around a KTN black hole. This is also important from the astrophysical point of view, as the first clue of the existence of GMM was reported \cite{Chakraborty-Bhattacharyya2018} in the astrophysical collapsed object GRO J1655-40 by using the X-ray observational data. It was also shown that GRO J1655-40 could be better described by the KTN spacetime instead of the Kerr spacetime \cite{cbgm2}. The warped accretion disk was observed in GRO J1655-40 \cite{martin} as well. Primordial black holes \cite{cb22} and M87* \cite{gcyl} could also contain GMM. Flattening out of the galaxy rotation curves may be a manifestation of the presence of GMM, without the need of dark matter particles \cite{gov, rug}. Thus, one should probe the inner accretion disk around a KTN black hole and find how the inner accretion disk structure is affected due to the presence of both the Kerr parameter and GMM. 

Lynden-Bell and Nouri-Zonoz \cite{lnbl} were perhaps the first to argue that the signatures of GMM might be found in the spectra of supernovae, quasars, or active galactic nuclei \cite{Chakraborty-Bhattacharyya2018}. However, the existence of GMM has been a subject of debate \cite{dihi} especially with regard to the closed timelike curve (CTC). Actually, Misner \cite{mis} wanted to present an entirely nonsingular cosmological model (homogeneous and anisotropic) with the Taub-NUT metric (described by the mass and GMM) which has a coordinate singularity at $\theta =\pi$ \cite{rs2}, known as the Misner string. To avoid this string singularity, Misner imposed the time periodicity condition \cite{rs2, cb22} which raises the causality violation issue due to the presence of CTC. Recently, it has been shown \cite{cle1, cle2} that the Taub-NUT spacetime is free from causal pathologies for freely falling observers if the time periodicity condition is not imposed \cite{hen, bor1}, and hence, some longstanding obstructions to accept the Taub-NUT solution as physically relevant are removed \cite{cle1, cb22}. In fact, it has recently been shown in \cite{cle1, cle2} that there are no closed timelike or
null geodesics in the
Taub-NUT spacetime, and, hence, the freely falling observers should not encounter causality violations \cite{wu22, wu23}. Moreover, as the collapsed object GRO J1655-40 was shown to be better described with the GMM \cite{Chakraborty-Bhattacharyya2018, cbgm2}, and due to of the availability of the state-of-the-art astrophysical observation facilities now a days, it could be more appropriate to take a practical approach toward understanding the nature of these kind of nonstandard astrophysical backgrounds \cite{dihi}.

Being motivated with all these, in this paper, we study the tilted thin accretion disk around a KTN black hole. In section \ref{s2}, we discuss the behavior of the LT effect in the KTN black hole. We describe the basic formalism of the tilted disk in section \ref{s3}. Our result for the radial profile of the tilted disk is described in section \ref{s4.0}. We summarize and conclude in section \ref{s5}.

\section{\label{s2}Lense-Thirring precession in Kerr-Taub-NUT spacetime}

The KTN metric is expressed here in 
Schwarzschild-like coordinates ($t,~R,~\th , ~ \phi$) in the geometrized{\footnote{The geometrized unit is considered only in section \ref{s2}, as it is easy to handle the general relativistic calculation in this unit. The cgs unit is used in the rest of the paper.} unit ($G=c=1$; where $G$ is the Newtonian gravitational constant and $c$ is the speed of light in vacuum) \cite{mill, Chakraborty-Bhattacharyya2018},
\begin{equation}
	ds^2=-\f{\d}{p^2}(dt-A d\phi)^2+\f{p^2}{\d}dR^2+p^2 d\th^2
	+\f{1}{p^2}\sin^2\th(adt-Bd\phi)^2,
	\label{metric}
\end{equation}
where
\begin{eqnarray}\nonumber
	\d&=&R^2-2MR+a^2-n^2,\\
	\nonumber p^2&=&R^2+(n+a\cos\th)^2,\\
	\nonumber A&=&a \sin^2\th-2n\cos\th,\\
 B&=&R^2+a^2+n^2,
\end{eqnarray}
with $M$ is the mass of the spacetime, $a$ is the Kerr parameter and $n$ is the NUT parameter. The outer horizon is located at $R_+=M+\sqrt{M^2+n^2-a^2}$, and the singularity is located at $\left[R=0, ~\th=\cos^{-1}(-n/a)\right]$ \cite{gcyl}.

The exact LT precession frequency ($\O_{\rm LT}^{\rm KTN}$) for the prograde orbits at $\th \ra \pi/2$ in the KTN spacetime is derived as \cite{Chakraborty-Bhattacharyya2018}
\begin{eqnarray}\nonumber
	\O_{\rm LT}^{\rm KTN} &=& \f{m^{1/2}}{R^{1/2}~(R^2+n^2) + a~m^{1/2}} \left[1-\f{1}{m^{1/2}~(R^2+n^2)}  \times
	\left[M(R^6-n^6+15n^4R^2-15n^2R^4) \right. \right.
	\\
	&& \left. \left. +2n^2 R (3R^4-2n^2R^2+3n^4) + 16M^2n^2R^3 - 4aR^{1/2}m^{1/2}(n^2+MR)(n^2+R^2) \right. \right. \nonumber
 \\
&& \left. \left.	-a^2\left\{M(n^4+6n^2R^2-3R^4)-8n^2R^3 \right\}\right]^{1/2}\right],
	\label{omega_LT1} 
\end{eqnarray}
where $m=M~(R^2-n^2)+2~n^2R$. For $n \ra 0$, eq. (\ref{omega_LT1}) reduces to \cite{ka}
\begin{eqnarray} 
\O_{\rm LT}= \f{M^{1/2}} {\left(R^{3/2}+aM^{1/2}\right)}  \times
 \left[1-\left(1-\f{4aM^{1/2}}{R^{3/2}}
 +\f{3a^2}{R^2} \right)^{1/2}\right], 
 \label{ltk}
\end{eqnarray}
which is well-known expression of the orbital plane precession frequency ($\O_{\rm LT}$) in the Kerr spacetime.

In this work, we study the tilted thin accretion disk around a KTN black hole of mass $M$ neglecting the higher order terms related to $a/R$ and $n/R$}. Thus, let us first write down the expression of $\O_{\rm LT}^{\rm KTN}$ (eq. \ref{omega_LT1}) upto the second order of $a_* (\equiv a/M)$ and $n_* (\equiv n/M)$, $i.e.$,
\begin{equation}
	\O_{\rm p} \equiv \O_{\rm LT}^{\rm KTN} \approx \frac{2a_* M^2}{R^3}-\frac{3a_*^2M^{5/2}}{2R^{7/2}}-\frac{2n_*^2M^{3/2}}{R^{5/2}} \times (1-2M/R)^2+ \mathcal{O}(a^3_*,n^3_*)
 \label{approx}.
\end{equation}
The reason for considering upto the second order of both the parameters in eq. (\ref{approx}) is that the lowest order of NUT parameter appears in the metric (eq. \ref{metric}), and thereby in the LT precession frequency expression (eq. \ref{omega_LT1}) is in the second order. Thus, in order to observe the effect of GMM in our study, it is necessary to take into account up to the second order of $n_*$. It is also evident from eq. (\ref{approx}) that no linear order term of $n_*$ appears in the expression of $\O_{\rm p}$. To compensate $n_*^2$ in eq. (\ref{approx}), we consider upto the second order term of Kerr parameter (i.e., $a_*^2$). The first term of eq. (\ref{approx}) represents the LT precession frequency for the Kerr black hole \cite{cap, fra1, fra2, mar, nel, nat, bcb19, li1, li2}. The second term of eq. (\ref{approx}) represents the precession due to the quadrupole moment of the Kerr black hole \cite{bp}. This second term is negligible compared to the first term and might be less important at the large distance (i.e., $R >> M$) as it varies $\sim a_*^2/R^{7/2}$. On the other hand, the third term which corresponds to the LT precession due to the NUT parameter/GMM, dominates over the second term as it varies $\sim n_*^2/R^{5/2}$. 
Note that, the modulus of third term can even dominate over the first term of eq. (\ref{approx}) depending on the numerical values of $a_*, ~n_*$ and $R$. This indicates that $\O_{\rm p}$ vanishes at a particular orbit of radius $R=R_0$, and the negative LT precession arises for $R > R_0$. The negative LT precession in the KTN black hole arises due to the sole effect of NUT charge \cite{Chakraborty-Bhattacharyya2018}, and it represents the orbital plane precession in the opposite direction \cite{ckp}. This peculiar behavior of the LT precession does not arise in case of the Kerr black hole where it follows the inverse cube law of distance ($\O_{\rm LT} \sim 2a_*M^2/R^3$). \footnote{The negative LT precession arises in the Kerr naked singularity for the range $R_{\rm ISCO}^{\rm Kerr} \leq R < 0.5625 a_*^2M$ \cite{ckp}. Thus, the value of $R_0$ in the Kerr spacetime is $R_0^{\rm Kerr}=0.5625 a_*^2M$. For the Kerr naked singularity, the negative LT precession arises in the orbit(s) of inner disk, whereas it arises in the outer portion ($R > R_0$) of the disk for the KTN black hole.}
In case of the KTN black hole, it is evident from eq. (\ref{approx}) that $\O_{\rm p}$ does not follow the same law for all combinations of $(a_*,n_*)$, and it can even vanish at $R_0$. The exact expression of $R_0$ can be obtained from eq. (\ref{approx}) by setting $\O_{\rm p}=0$ at $R=R_0$. However, the approximate form of $R_0$ can be expressed as,
\begin{eqnarray}
 R_0 \approx \f{M}{4n_*^4}\left(2 a_*^2  - 3 a_*^2 n_*^2 + 16 n_*^4 + 
 2a_* \sqrt{a_*^2  - 3 a_*^2  n_*^2 + 16n_*^4}\right)
 \label{R0}
\end{eqnarray}
by neglecting $\sim n_*^2M^{7/2}/R^{9/2}$ from the last term of eq. (\ref{approx}) and solve it. Note that the exact numerical value of $R_0$ is very close to the approximate value obtained from eq. (\ref{R0}), if $R_0$ really occurs close to the order of $\sim M$. In most of the cases, the value of $R_0$ is sufficiently greater than $M$. Thus, $R_0$ gives correct values in those cases\footnote{The exact value of $R_0$ is not important in this paper. We introduce this only to show how the tilting of the disk depends on the behavior of LT effect.}.

\begin{figure}[h]
	\begin{center}
		\includegraphics[width=0.7\columnwidth,height=0.8\columnwidth]{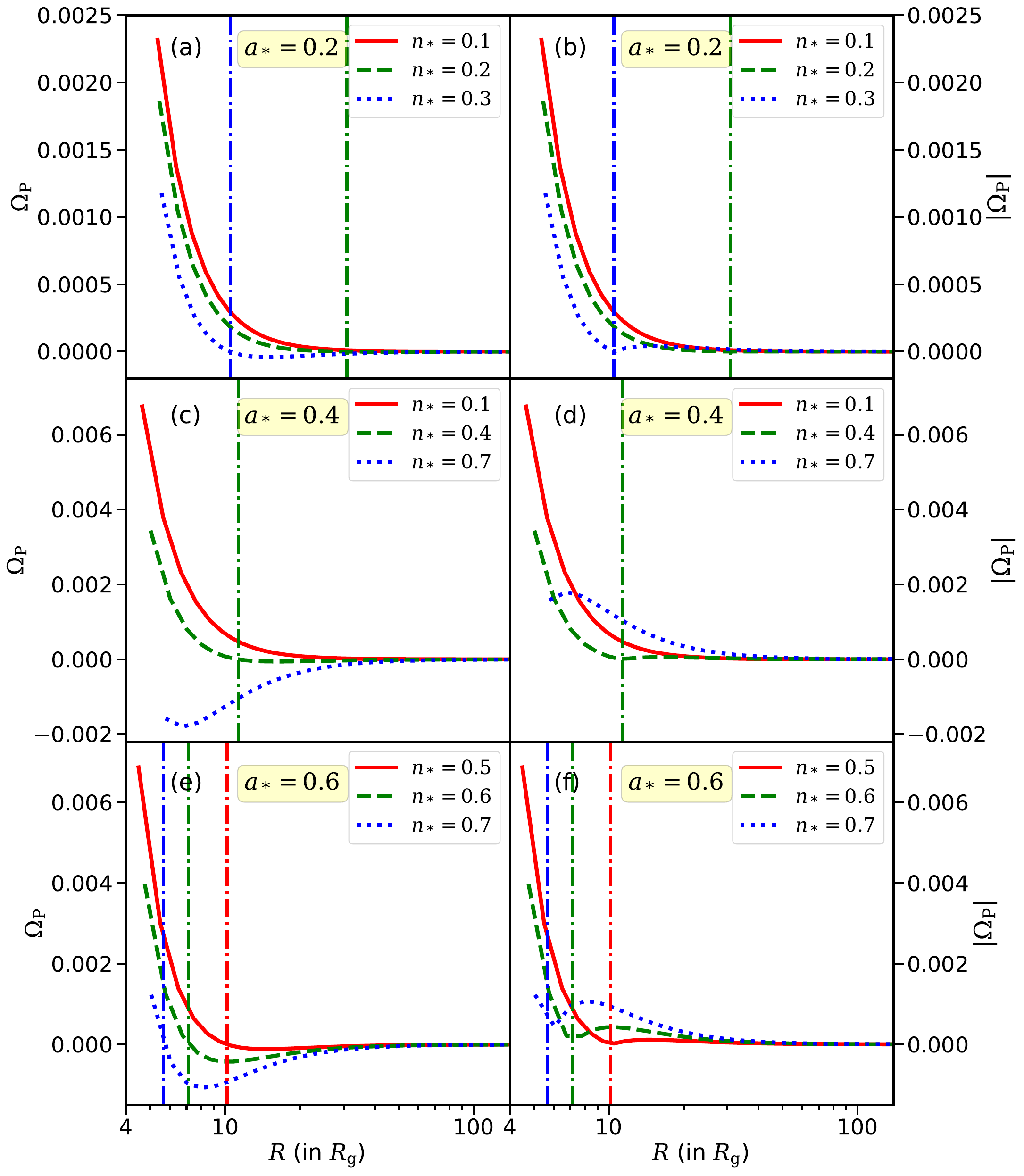}
	
	\end{center}
	\caption{Variation of $\O_{\rm p}$ (left panel)  and $|\O_{\rm p}|$ (right panel) with radial distance ($R$) for three different values of $n_*$ with $a_*=0.2$ (Panels (a) and (b)), $a_*=0.4$ (Panels (c) and (d)), and $a_*=0.6$ (Panels (e) and (f)). All the corresponding $R_0$ are represented in the figure by the dot-dashed lines with respective colors of each set of ($a_*,n_*$). The values of $R_0$ for solid red curves for Panels (a)-(d) are greater than the value considered in this figure along the X-axes, and hence the corresponding dot-dashed red lines are not seen for those said panels. On the other hand, as $R_0 < R_{\rm ISCO}$, the blue dot-dashed line is not shown in Panels (c)-(d). Note that $\O_{\rm p}$ and $|\O_{\rm p}|$ tend to zero for $R \rightarrow \infty$, as expected.  See \S~ \ref{s2} for details.}
	\label{omega_0p6_0p4_0p2}
	
\end{figure}
In accordance with eq. (\ref{approx}), the LT effect for a KTN black hole exhibits dominance either by the Kerr parameter ($a_*$) or the NUT parameter/GMM ($n_*$) at any given radial coordinate, as there is a competition between these parameters at $R$. To investigate the characteristic behaviors of $\O_{\rm p}$ and $|\O_{\rm p}|$, we explore three distinct sets of Kerr parameters ($a_*$) in figure \ref{omega_0p6_0p4_0p2}, each followed by three different values of the NUT parameter ($n_*$). This will help us for the comprehensive understanding of the influence of these parameters on the LT frequency and tilted angle of the KTN black hole at a later stage. It is useful to note here that all the curves of figure \ref{omega_0p6_0p4_0p2} are started from their respective innermost stable circular orbits (ISCOs). The numerical value of the ISCO for a specific combination of $(a_*, n_*)$ can be obtained by solving the following ISCO equation for prograde orbit \cite{cbgm2, Chakraborty-Bhattacharyya2018}
\begin{eqnarray}\nonumber
	&& M(R^6-n^6+15n^4R^2-15n^2R^4)-2M^2 R (3R^4-2n^2R^2+3n^4)-16n^4R^3 
	\\  
	& + & 8aR^{3/2}m^{3/2} +a^2\left\{M(n^4+6n^2R^2-3R^4)
	-8n^2R^3 \right\}=0
	\label{isco_ktn}.
\end{eqnarray}
However, as the LT torque mainly depends on the value of $\O_{\rm p}$, we present the corresponding analyses in figure \ref{omega_0p6_0p4_0p2}. In Panels (a) and (b) of figure \ref{omega_0p6_0p4_0p2}, we examine the variations of $\O_{\rm p}$ and $|\O_{\rm p}|$, where the Kerr parameter is set to $a_*=0.2$ with NUT parameter values as $n_*=0.1$ (red solid line), $n_*=0.2$ (green dashed line), and $n_*=0.3$ (blue dotted line). The values of $R_0$ for $(a_*,n_*)=(0.2,0.1), (0.2,0.2)$, and $(0.2,0.3)$ are $402 R_{g}, 31 R_{g}$, and  $10.5 R_{g}$, respectively. Similarly, Panels (c) and (d) of figure \ref{omega_0p6_0p4_0p2} show the same variations with $a_*=0.4$, paired with $n_*=0.1$ (red solid line), $n_*=0.4$ (green dashed line), and $n_*=0.7$ (blue dotted line). The values of $R_0$ for $(a_*,n_*)=(0.4,0.1), (0.4,0.4)$, and $(0.4,0.7)$ are $1583 R_{g}, 11.3 R_{g}$, and  $4.7 R_{g}$, respectively. Panels (e) and (f) of figure \ref{omega_0p6_0p4_0p2} are illustrated for $a_*=0.6$, along with $n_*=0.5$ (red solid line), $n_*=0.6$ (green dashed line), and $n_*=0.7$ (blue dotted line). The values of $R_0$ for $(a_*,n_*)=(0.6,0.5), (0.6,0.6)$, and $(0.6,0.7)$ are $10.2R_{g}, 7.1R_{g}$, and  $5.7R_{g}$, respectively.  Note that all the corresponding $R_0$ are represented in figure \ref{omega_0p6_0p4_0p2} by the dot-dashed lines
with the respective color of each set of $(a_*,n_*)$. As the X-axis in each panel of figure \ref{omega_0p6_0p4_0p2} is considered up to $\sim 100R_g$, one cannot see those $R_0$ which occur at $R_0 > 100R_g$. For example, the dot-dashed red lines are not seen in Panels (a)-(d) of figure \ref{omega_0p6_0p4_0p2}. For $(a_*, n_*)=(0.4,0.7)$, $R_0~(=4.7 R_g)$ occurs inside the ISCO radius $(R_{\rm ISCO}=5.8 R_g$), i.e., $R_0 < R_{\rm ISCO}$. Thus, the blue
dot-dashed line is not shown in Panels (c)-(d), as it is not feasible. For this particular case, $\Omega_{\rm p}$ always has a negative value for $R \geq R_{\rm ISCO}$. These six plots in figure \ref{omega_0p6_0p4_0p2} provide a clear depiction of the interplay between GMM and the Kerr parameter on the value of $\O_{\rm p}$. The behavior of all the curves of $|\O_{\rm p}(R)|$ look almost similar. If the value of $R$ decreases from the outer orbit to the ISCO, $|\O_{\rm p}(R)|$ first increases, attains a peak, then decreases to zero, and increases again depending on the location of the ISCO.
   Comparing   three sets of figures, we infer that, $\O_{\rm p}$ or 
   $|\O_{\rm p}|$ do not follow the same pattern. They depend on the combinations of $a_*$ and $n_*$. Thus, the behavior of the tilted angle of the accretion disk is subject to dependencies on the values of $a_*$ and $n_*$ along with the other parameters, which we discuss as we proceed.

\section{\label{s3}Formalism: Tilted and warped disk equation}
In our investigation, we focus on a spinning KTN black hole characterized by a small NUT parameter/GMM at its core. This scenario assumes a Keplerian disk configuration with an aspect ratio $H/R \ll 1$, where $H$ represents the disk thickness and $R$ signifies the radial distance from the black hole. Notably, the black hole's spin axis aligns with the $z$ axis and the accretion disk exhibits a tilt concerning the black hole's spin axis. The disk consists of the circular rings with width $\Delta R$ and surface density $\Sigma(R,t)$. The angular momentum per unit surface area on each annulus of the disk is defined as,  $\textbf{L}(R,t)=\Sigma R^2 \O(R)\textbf{l}(R,t)$ , where $\textbf{l}$ is the unit tilt vector directed normal to the plane of the disk and $\O(R)$ is the Keplerian angular speed. We adopt the assumption of a small tilt angle following \cite{sf}, implying that $\textbf{l}$ can be approximated as $(l_{x}, l_{y}, 1)$.
Moreover, we consider the accretion disk to be sufficiently viscous, satisfying the condition $\alpha > H/R$ (where, $\a$ is the Shakura–Sunyaev parameter).
In this viscous regime, warping is transported
diffusively in the disk \cite{Papaloizou-Pringle1983}. In the opposite regime, i.e., $\alpha < H/R$, which is not  considered in this work, the warping disturbances propagate in a
wave-like manner \cite{Ivanov-Illarionov1997, Lubow-etal2002}. In the latter case, people have
also been very interested in the radial tilt oscillations which could be developed in the inner disk around
a Kerr black hole \cite{Ivanov-Illarionov1997, di97, Lubow-etal2002}.

In this paper, our focus lies in investigating the effect of both the viscous and LT torques within the disk in a steady state. Thus, following Pringle's \cite{Pringle1992}
equation with LT precession, one can rewrite the basic tilted/warped disk equation (equation 2 of \cite{sf}) as

\begin{equation}
	\frac{1}{R}\frac{\p}{\p R}\left[\left(\frac{3R}{L}\frac{\p}{\p R}\left(\nu_1 L\right)-\frac{3}{2}\nu_1 \right)\textbf{L}+\frac{1}{2}\nu_2 R L \frac{\p \textbf{l}}{\p R}\right]+(\boldsymbol{\O}_{\rm p}\times\textbf{L})=0.
	\label{angular_steady_state}
\end{equation}
In the context of accretion disk dynamics, $\nu_1$ of eq. (\ref{angular_steady_state}) represents the viscosity associated with the azimuthal shear, pertaining to the $(R,\phi)$ component of shear, while $\nu_2$ denotes the viscosity linked with the vertical shear, corresponding to the $(R,z)$ component of shear \cite{Papaloizou-Pringle1983}.  
The ratio $\nu_2/\nu_1$, termed as viscous anisotropy, holds significance in this context and can be associated with $\alpha$ for small amplitude warps, as evidenced in prior investigations \cite{Ogilvie-1999}. Note that this is not 
 done according to the method described in \cite{mar}, where $\nu_1$ and $\nu_2$ were considered as a function of $R$. In this paper, we assume both of these viscosities are independent of the radial distance following \cite{sf, cb}. The ratio of $\nu_2$ and $\nu_1$ is expressed as \cite{Ogilvie-1999}
\begin{equation}
	\frac{\nu_2}{\nu_1}=\frac{1}{2\alpha^2}.\frac{4(1+7\alpha^2)}{4+\alpha^2}.
	\label{anositropy}
\end{equation} 
To account for the relativistic effects induced by the presence of a KTN black hole, we extend eq. (\ref{angular_steady_state}) to incorporate the external torque arising from the LT precession.
The external torque ($\boldsymbol{\tau}$) in eq. (\ref{angular_steady_state}) due to the LT precession is given by,
\begin{equation}
	\boldsymbol{\tau}=\boldsymbol{\O}_{\rm p} \times\textbf{L},
	\label{LT}
\end{equation}
where $\O_{\rm p}$ in the cgs unit is written as (see eq. \ref{approx}),
\begin{eqnarray}
	\O_{\rm p}&=&\frac{c}{R}\left(\frac{2a_*R_g^2}{R^2}-\frac{3a_*^2 R_g^{5/2}}{2R^{5/2}}-\frac{2n_*^2R_g^{3/2}}{R^{3/2}}(1-2R_g/R)^2\right),
	\label{omega_LT}
\end{eqnarray}
with $R_g~(\equiv GM/c^2)$ as the gravitational radius.

Taking the scalar product of $\textbf{l}$ \cite{sf} with eq. (\ref{angular_steady_state}), we get 
\begin{equation}
	\frac{1}{R}\frac{\p}{\p R}\left[R\frac{\p}{\p R}\left(\nu_1 L\right)-\frac{1}{2}\nu_1 L\right]=0,
\end{equation}
under the small tilt angle approximation (we have ignored the term $\vert\p \textbf{l}/\p R\vert^2$). Solving the above equation, one obtains \cite{sf}, 
\begin{equation}
	L(R)=C_2 R^{1/2}-2 C_1,
	\label{L1_L2}
\end{equation}
where $C_1$ and $C_2$  are the integration constants. They are obtained as 
$C_2=\sqrt{GM}\Sigma_{\infty}$ and
$C_1=\frac{1}{2}\sqrt{GMR_{\rm in}}\left(\Sigma_{\infty}-\Sigma_{\rm in}\right)$ \cite{cb},
where $\Sigma \rightarrow \Sigma_{\infty}$ at $R \rightarrow \infty$ and $\Sigma \rightarrow \Sigma_{\rm in}$ at $R \rightarrow R_{\rm ISCO}$. Here, $R_{\rm in}$ corresponds to the inner edge radius of the disk, that is, in fact, the ISCO radius ($R_{\rm in} \equiv R_{\rm ISCO}$) for a KTN black hole. Note that one can obtain $R_{\rm ISCO}$ by solving eq. (\ref{isco_ktn}). Finally, substituting the above expressions for $C_1$ and $C_2$ into eq. (\ref{L1_L2}), we obtain the expression for $L(R)$ in steady state as, 
\begin{equation}
	L(R)=\sqrt{GM}\left[R^{1/2}\Sigma_{\infty}+R^{1/2}_{\rm in}\left(\Sigma_{\rm in}-\Sigma_{\infty}\right)\right].
	\label{L_1}
\end{equation}

The steady state distribution of the surface density can similarly be obtained from eq. (\ref{L_1}) as
\begin{equation}\label{sig}
	\Sigma(R)=\Sigma_{\infty}+\left(R_{\rm in}/R\right)^{\frac{1}{2}}\left(\Sigma_{\rm in}-\Sigma_{\infty}\right).
\end{equation}

Now, substituting $L(R)$ in eq. (\ref{angular_steady_state}) we obtain, 
\begin{equation}
	\frac{1}{R}\frac{\p}{\p R}\left[ 3\nu_1 C_1 \textbf{l}+\frac{1}{2}\nu_2 R L \frac{\p \textbf{l}}{\p R}\right]+(\boldsymbol{\O}_{\rm p}\times\textbf{L})=0.
	\label{angular_steady_state2}
\end{equation}
Equation (\ref{angular_steady_state2}) can be decomposed into $x$ and $y$ components (i.e., $l_x$ and $l_y$) of the tilt vector $(\textbf{l})$ as
\begin{equation}
	\frac{\p}{\p R}\left(3\nu_1 C_1 l_x+\frac{1}{2}\nu_2 R L \frac{\p l_x}{\p R}\right)=\omega_{\rm p} L l_y,
	\label{lx}
\end{equation} 

and 
\begin{equation}
	\frac{\p}{\p R}\left(3\nu_1 C_1 l_y+\frac{1}{2}\nu_2 R L \frac{\p l_y}{\p R}\right)=-\omega_{\rm p} L l_x,
	\label{ly}
\end{equation}
where $\boldsymbol{\omega}_{\rm p} \times\textbf{l}=\left(-\omega_{\rm p}  l_y,\omega_{\rm p}  l_x,0\right)$ and $\omega_{\rm p} = c\left(\frac{2a_*R_g^2}{R^2}-\frac{3a_*^2 R_g^{5/2}}{2R^{5/2}}-\frac{2n_*^2R_g^{3/2}}{R^{3/2}}(1-2R_g/R)^2\right).$  \\
Combining eqs. (\ref{lx}) and (\ref{ly}), we obtain
\begin{equation}
	\frac{\p}{\p R}\left(3\nu_1 C_1 W+\frac{1}{2}\nu_2 R L \frac{\p W}{\p R}\right)=-i\omega_{\rm p}  L W,
	\label{W}
\end{equation} 
where we define $W=l_x+i l_y=\beta e^{i\gamma}$ with $\beta=\sqrt{l_x^2+l_y^2}$ is the tilt angle and $\gamma=\tan^{-1}\left(l_y/l_x\right)$ is the twist angle. eqs. (\ref{lx}) and (\ref{ly}) refer the evolution and characteristics of the warped disk around a KTN black hole in the steady state. 

Here, we use the dimensionless form of some parameters for our convenience in the mathematical calculations. For instance, the dimensionless form for $L$ \cite{Banerjee-etal2019} is given by
$L\rightarrow L/C_1=(C\sqrt{R}-2)$, where
$C=\frac{2z_{\rm in}}{z_{\rm in}-1}\frac{1}{\sqrt{R_{\rm in}}}$ and $z_{\rm in}=1+\frac{2 C_1}{L(R_{\rm in})}=\frac{\Sigma_{\infty}}{\Sigma_{\rm in}}$ \cite{Banerjee-etal2019}. We make $R$ dimensionless by replacing $R \ra \frac{R}{R_g}$, $\xi \ra \frac{cR_g}{\nu_2}$ and $\eta \ra \frac{6\nu_1}{\nu_2}$. 
Using the above-mentioned scheme and by following \cite{Banerjee-etal2019}, we obtain the dimensionless form of eqs. (\ref{lx}) and (\ref{ly}) as
\begin{equation}
	R\frac{\p^2 l_x}{\p R^2}+\left[(\eta+1)\frac{C_1}{L}+3/2\right]\frac{\p l_x}{\p R}=\frac{2\omega^{\rm KTN}_{\rm LT} l_y}{\nu_2}=2\xi\overline{\omega} l_y,
	\label{lxf}
\end{equation}
and
\begin{equation}
	R\frac{\p^2 l_y}{\p R^2}+\left[(\eta+1)\frac{C_1}{L}+3/2\right]\frac{\p l_y}{\p R}=-\frac{2\omega^{\rm KTN}_{\rm LT} l_x}{\nu_2}=-2\xi\overline{\omega} l_x .
	\label{lyf}
\end{equation}
where,
\begin{equation}
 \overline{\omega} = \left(\frac{2a_*}{R^2}-\frac{3a_*^2}{2R^{5/2}}-\frac{2n_*^2}{R^{3/2}}(1-2/R)^2\right).
\end{equation}

Now, to see the behavior of the tilted accretion disk, we have to solve the eqs. (\ref{lxf}) and (\ref{lyf}) numerically with proper boundary conditions. We solve these equations as a boundary value problem. For this case, we fix the boundary at $R_f$ as an outer edge of the accretion disk and $R_{\rm in}$ as the inner edge of the disk. Thereafter, we also fix the tilted and twisted angle at the outer and inner edge. For a set of fixed parameters ($i.e.,$ $a_*$, $n_*$, $M$, $\nu_2$ and $\eta$) we solve these above-mentioned equations bounded by $R_f$ and $R_{\rm in}$ and see the behaviour of $\beta$ with radial coordinate ($R$).
To solve the before-mentioned warped disk equations (eqs. \ref{lxf} and \ref{lyf}), we need four boundary conditions, which are taken as \cite{Banerjee-etal2019}
\begin{eqnarray}
	l_x(R_{\rm in})&=&\beta_i \cos(\gamma_i),\ l_y(R_{\rm in})=\beta_i \sin(\gamma_i),
	\label{b1}
\end{eqnarray}
and, 
\begin{eqnarray}
	l_x(R_{f})&=&\beta_f, \ l_y(R_{f})=0.
	\label{b2}
\end{eqnarray}
We define several key parameters:  $\gamma_i$, $\beta_i$, and $\beta_f$, representing the twist angle at $R_{\rm in}$, tilt angle at $R_{\rm in}$, and tilt angle at $R_f$ of the disk, respectively. At the outer edge of the disk, we assume the twist angle to be zero since the effect of LT precession is negligible there \cite{Banerjee-etal2019}. Consequently, $l_y$ can be assumed to be zero at the outer edge. Note that \cite{cb} focused on the disk inner edge tilt, and thus assumed the inner edge twist to be zero. Here, we retain the disk inner edge twist term to make our solutions more general following \cite{Banerjee-etal2019}. However, choosing different numerical value for $\gamma_i$, the tilt profile does not show any significant differences qualitatively and quantitatively. 
The inner edge of the disk ($R_{\rm in}$) is considered here as the ISCO radius ($R_{\rm ISCO}$) for a prograde disk. We have shown the nature of a tilted disk considering non-zero tilt angle ($\beta_i \ne 0$) and zero tilt angle ($\beta_i = 0$) at $R_{\rm ISCO}$.

\section{\label{s4.0}Results and discussion}
   
 In this section, we investigate the behavior of the tilt angle radial profile $(\beta(R))$ as a function of the parameters $a_*$, $n_*$, $\beta_i$, $\nu_2$, and $\eta$ in detail. In case of the Kerr black hole, it was earlier shown \cite{Banerjee-etal2019} that the interplay between the LT torque (controlled by $M, a_*, \beta_i$) and viscous torque (controlled by $\nu_2$) in the plane of the disk decides the radial profile of $\beta$. In case of the KTN black hole, the LT torque is additionally controlled by $n_*$ as well. In section \ref{s2}, we have shown that the LT effect (eq. \ref{approx}) can decrease or increase or even vanish at a particular orbit ($R_0$) depending on the values of $a_*$ and $n_*$. The similar effect does not arise for the Kerr black hole. In case of the KTN black hole, although there is certainly an interplay between the LT torque and viscous torque, one cannot neglect another interplay between $a_*$ and $n_*$ inside the LT torque. Due to the interplay between three primary parameters ($a_*, n_*$ and $\nu_2$), some peculiar effect will appear around $R_0$ (not exactly at $R_0$), which we discuss as we proceed.
 
 \begin{figure}[h]
	\begin{center}
		\includegraphics[width=\columnwidth]{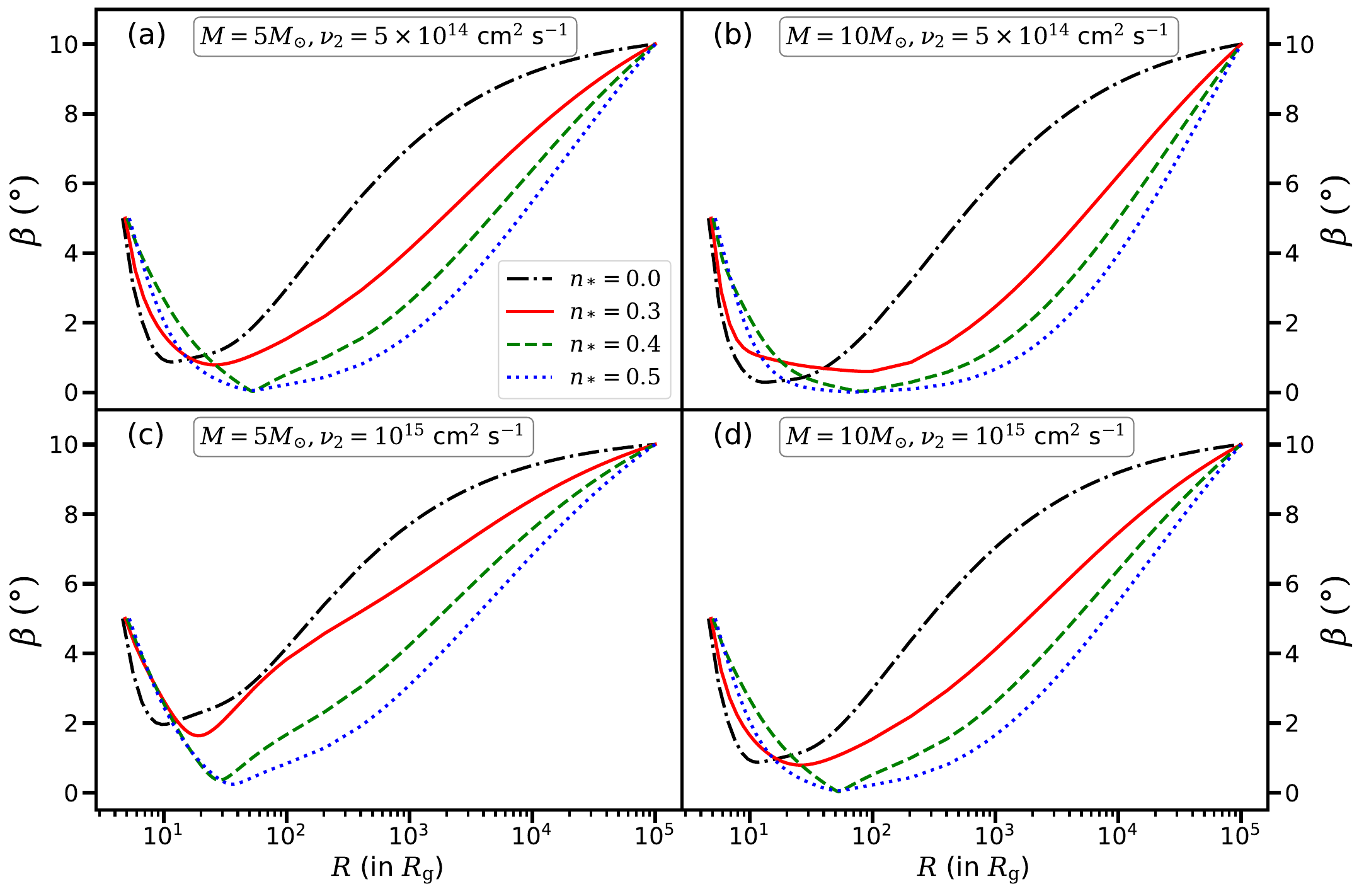}
	\end{center}
	\caption{Variation of tilt angle ($\beta$) with radial distance ($R$) for four different values of $n_*$ with a fixed value of $a_*=0.4, \beta_i=5^{\circ}, \eta=0.25$ and $z_{\rm in}=0.75$. The value of $M$ ($\nu_2$) is fixed in the plots of 1st/2nd column (row) with different values of $\nu_2$ ($M$) as mentioned in the inset. All the curves are started from $R_{\rm in} \equiv R_{\rm ISCO}(a_*,n_*)$ which are calculated using eq. (\ref{isco_ktn}). See \S~ \ref{s4.2} for details.}
	\label{beta_m_nu2}
\end{figure}

\subsection{\label{s4.1}Parameter values}
In order to discuss the behavior of $\beta(R)$, we need to choose suitable numerical values for the different parameters relevant to the astrophysical scenario. As we are mainly interested in the Galactic accreting black holes, we choose $M \sim 5-25 M_{\odot}$ (where $M_{\odot}$ is the solar mass) for most of the cases. The value of $\nu_2$ is considered as $10^{14}-10^{15}$ $\rm cm^2$ $\rm s^{-1}$ \cite{Frank}. $\eta$ is chosen as $\eta = 0.25$ which translates $\alpha = 0.156$ \cite{King}. We consider $z_{\rm in} =0.75$, as in our formalism $\Sigma_{\rm in}>\Sigma_{\infty}$. The inner edge twist and outer edge tilt to $\sim 5^\circ$ and $\sim 10^\circ$, respectively, throughout the paper following \cite{Banerjee-etal2019}. We consider $\beta_i$ as a free parameter following \cite{cb22, Banerjee-etal2019}, and use the range $0^\circ-10^\circ$ for the purpose of demonstration. Note that  we consider only the case of prograde rotation $(a_* > 0)$ in this paper,  and the numerical values of $a_*$ and $n_*$ are considered upto $0.7$ \cite{cap, fra1, fra2, mar, nel, nat, li1, li2, bcb19, Banerjee-etal2019}.

\subsection{\label{s4.2}Numerically Computed Radial Profiles of the Disk Tilt Angle}

 Figure \ref{beta_m_nu2} is plotted for $a_*=0.4$ with the four different values of $n_*$ : $n_* = 0.0$ (black dot-dashed solid line), $n_* = 0.3$ (red solid line), $n_* = 0.4$ (green dashed line), and $n_* = 0.5$ (blue dotted line). The values of $R_0$ are $\sim 24.5R_g, 11.3R_g$ and $7.3R_g$ for the combinations of $(a_*, n_*)\equiv(0.4,0.3), (0.4, 0.4)$ and ($0.4, 0.5$), respectively. $R_0$ is not feasible for the combinations of $(a_*, n_*)\equiv(0.4,0.0)$
 which means $\Omega_p$ cannot be zero in between $R_f$ and $R_{\rm ISCO}$. However, when transitioning from  the left to right $\Big[(a\rightarrow b), (c\rightarrow d)\Big]$, we maintain a constant value for $\nu_2$ while varying $M$ from $5M_{\odot}$ to $10M_{\odot}$. Comparing panels (a) and (b), or, panels (c) and (d), one can see that the transition point of disk misalignment moves to the larger $R$, as the mass increases. This is because, when the mass of the black hole increases, the LT torque increases, and the disk tries to align from a larger radius. Conversely, moving from the upper to the lower arrangement $\Big[(a\rightarrow c), (b\rightarrow d)\Big]$, we keep the mass $M$ fixed and adjust the viscosity parameter $\nu_2$ from $5\times10^{14}$ cm$^2$ s$^{-1}$ to $10^{15}$ cm$^2$ s$^{-1}$, with all other parameters held constant. 
 Comparing panels (a) and (c), or, panels (b) and (d), one can see that the transition point of disk misalignment moves to the smaller $R$, as the viscosity increases. As in panel (c), we increase viscosity more than in panel (a), keeping all other parameters fixed, the viscous torque increases, which is held out the disk against being aligned.

\begin{figure}[h!]
	\includegraphics[width=0.49\textwidth]{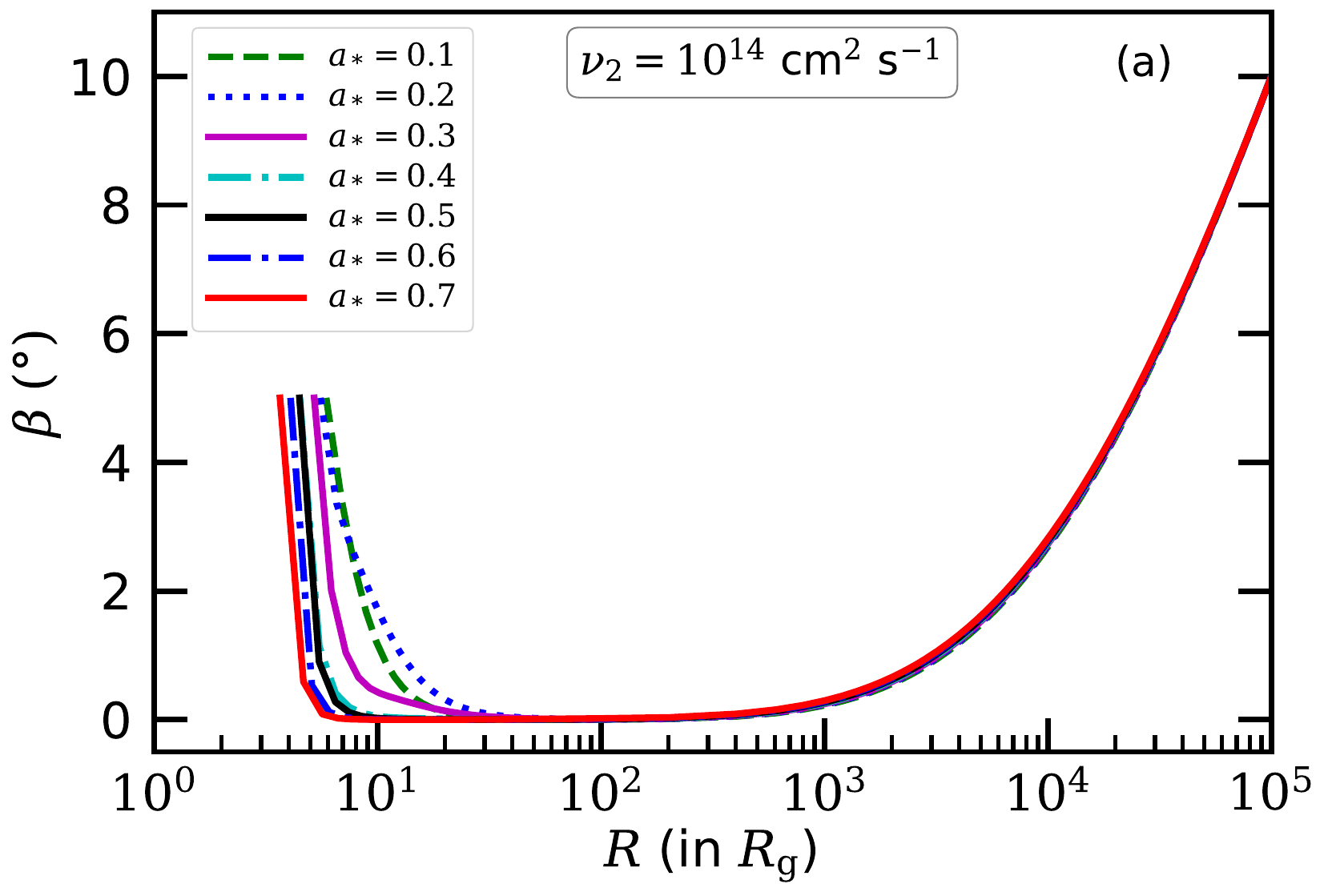}
	\includegraphics[width=0.49\textwidth]{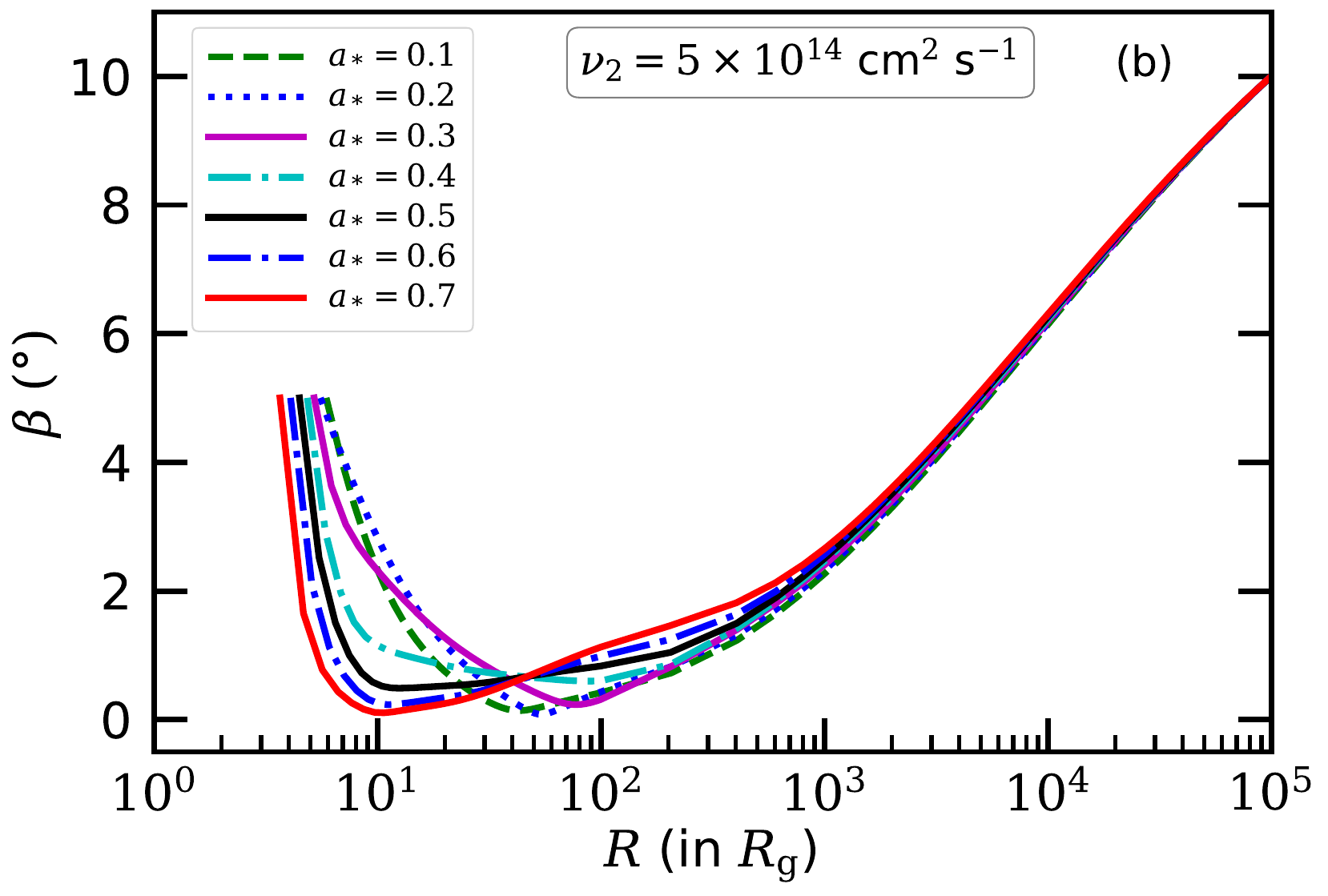}
	\includegraphics[width=0.49\textwidth]{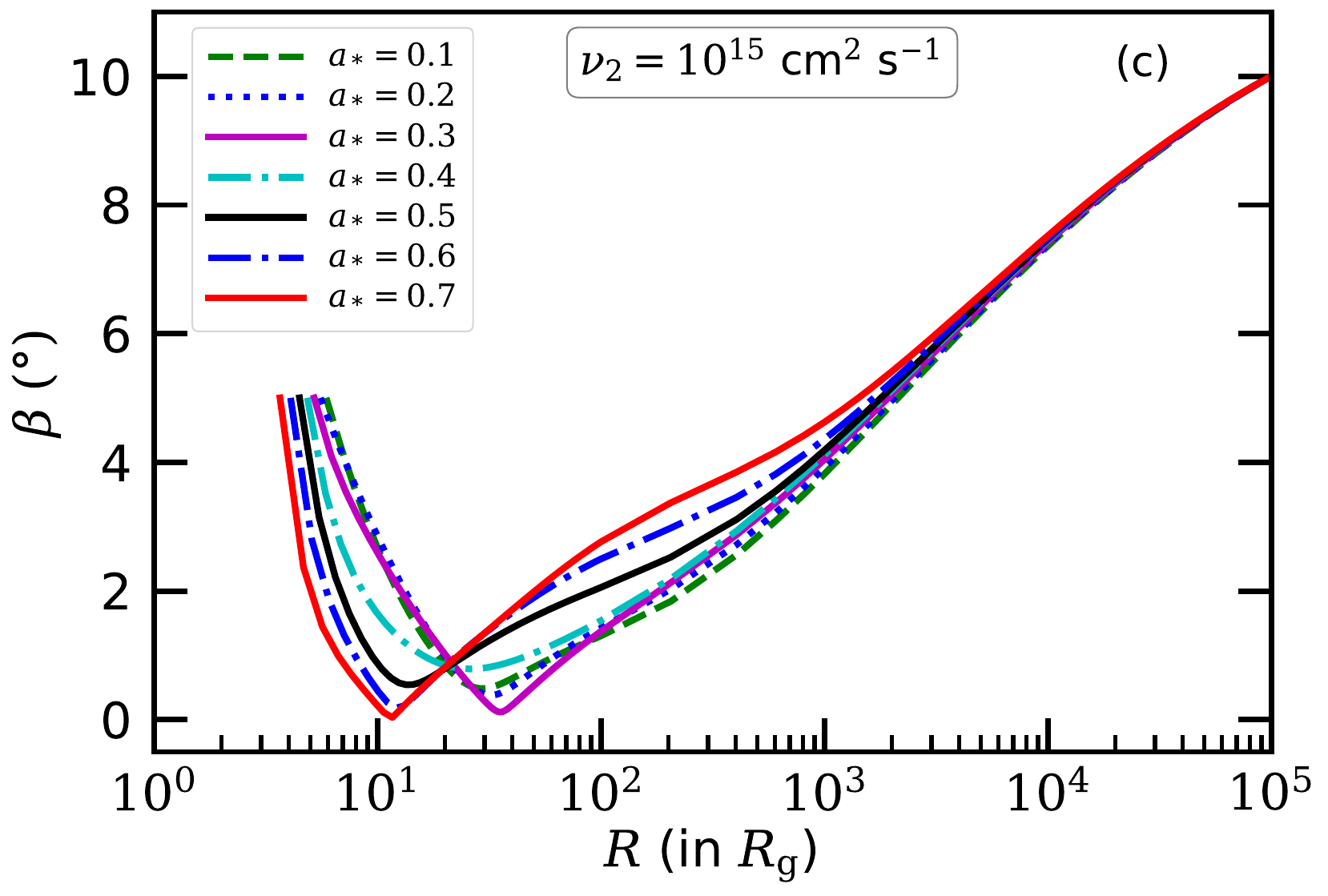}
	\includegraphics[width=0.49\textwidth]{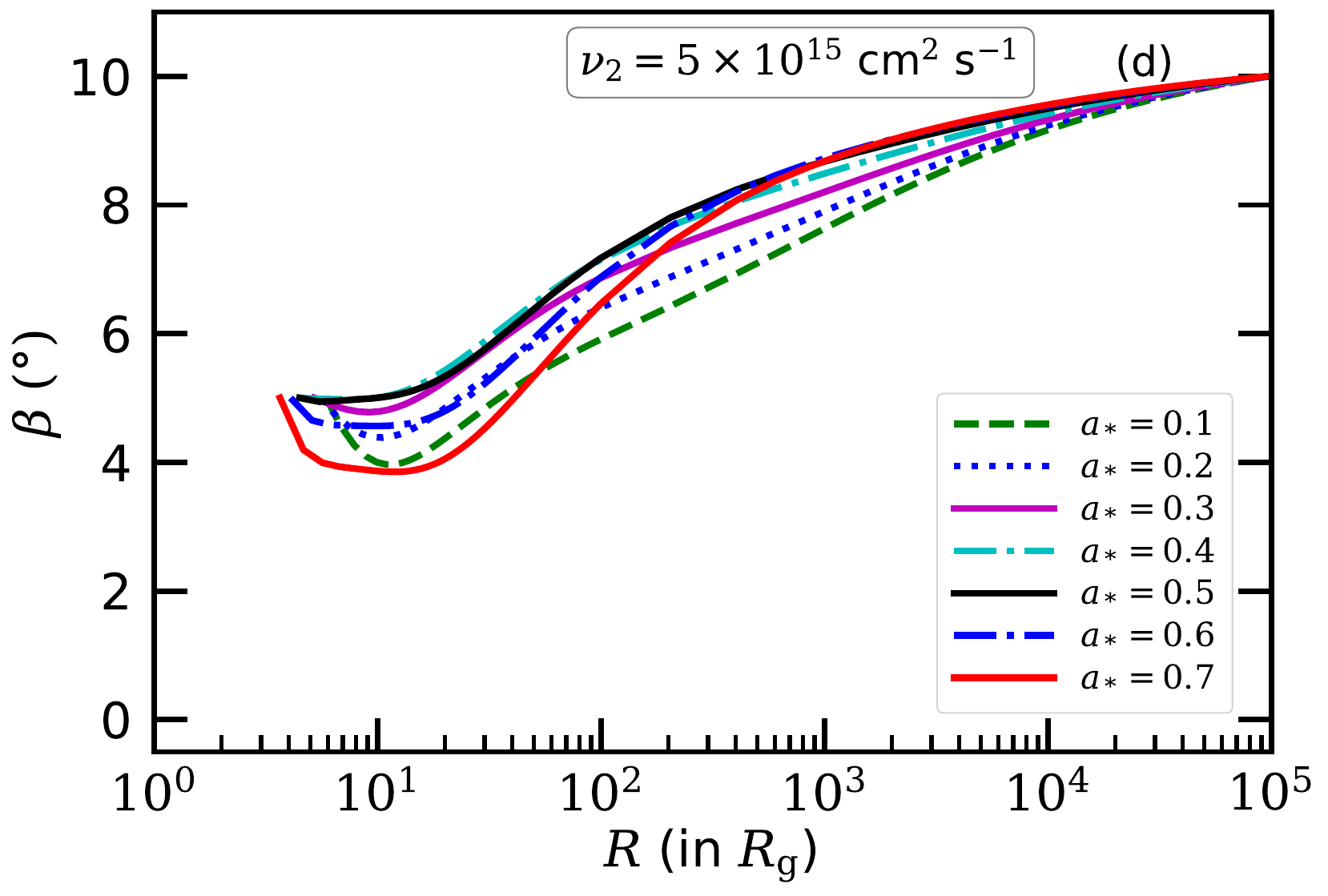}
	\caption{Variation of $\beta$ with $R$ for the different values of $a_*$ with a fixed value of $M=10M_{\odot}, \beta_i=5^{\circ}, n_*=0.3,\eta=0.25$ and $z_{\rm in}=0.75$. The value of $\nu_2$ changes in the plots as mentioned in the inset. All the curves are started from $R_{\rm in} \equiv R_{\rm ISCO}(a_*,n_*)$ which are calculated using eq. (\ref{isco_ktn}). See \S~ \ref{s4.2} for details.}
	\label{beta_r_nu_n0p3}
\end{figure}

Figure \ref{beta_r_nu_n0p3} illustrates the variation of $\beta$ with respect to the radial coordinate for different values of $\nu_2$. The upper-left figure corresponds to $\nu_2=10^{14}$ cm$^2$ s$^{-1}$, the upper-right figure to $\nu_2=5\times10^{14}$ cm$^2$ s$^{-1}$, the lower-left figure to $\nu_2=10^{15}$ cm$^2$ s$^{-1}$, and the lower-right figure to $\nu_2=5\times10^{15}$ cm$^2$ s$^{-1}$. The various curves within each figure represent different values of the Kerr parameter ($a_*$), ranging from $0.0$ to $0.8$ with an interval of $\Delta a_* =0.1$. It is notable that higher values of $a_*$ tend to align the angular momentum of the disk along the black hole's spin axis, resulting in smaller values of $\beta$. Observing the plots, it becomes evident that as the value of $\nu_2$ increases, the viscous torque begins to dominate over the LT torque for a fixed set of $(a_*,n_*)$. The competition between viscous and LT torques plays a crucial role in determining the orientation of the accretion disk, with higher values of $\nu_2$ exerting a greater influence on the disk's inclination relative to the black hole's spin axis.

\begin{figure}[h!]
	\includegraphics[width=0.49\textwidth]{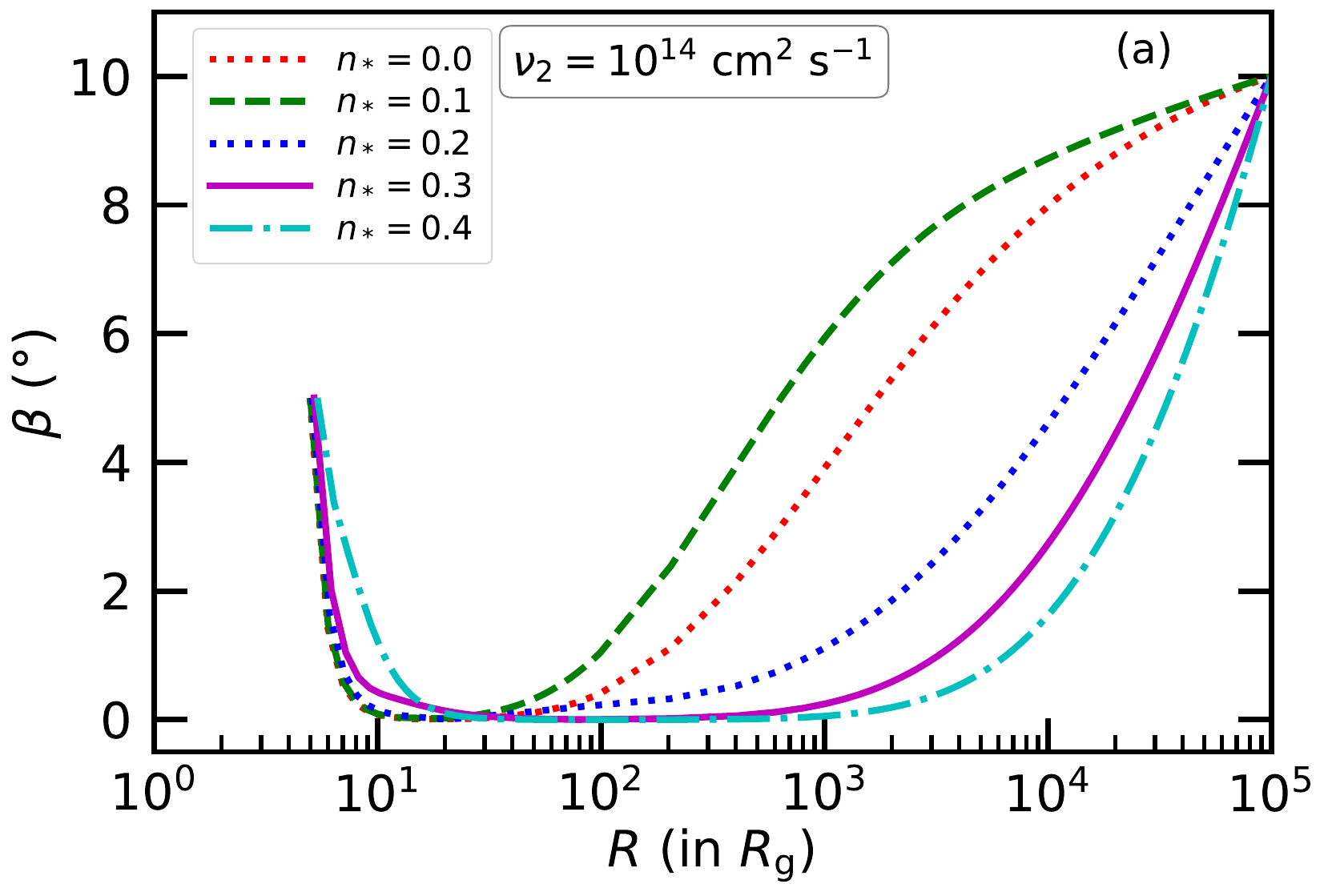}
	\includegraphics[width=0.49\textwidth]{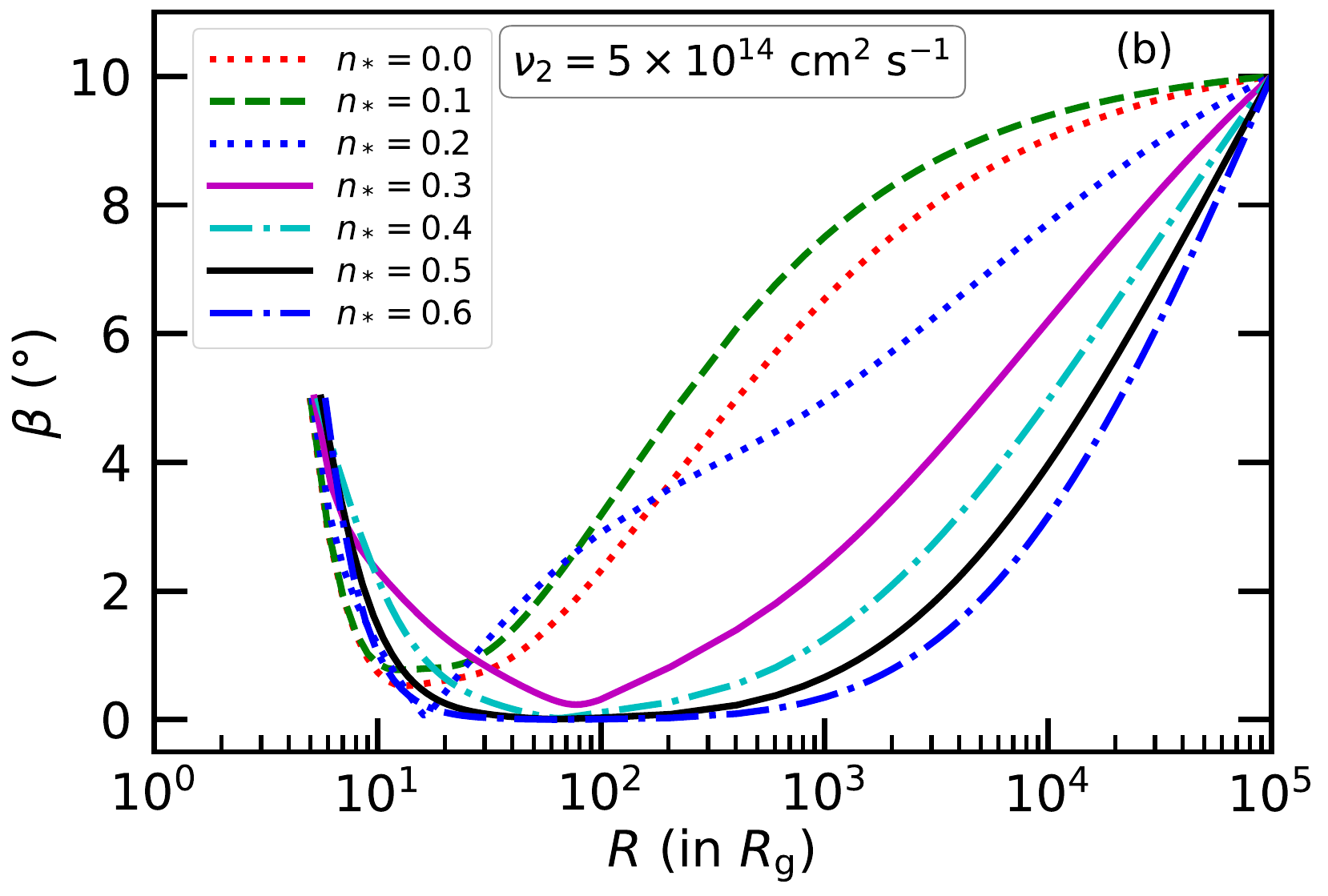}
	\includegraphics[width=0.49\textwidth]{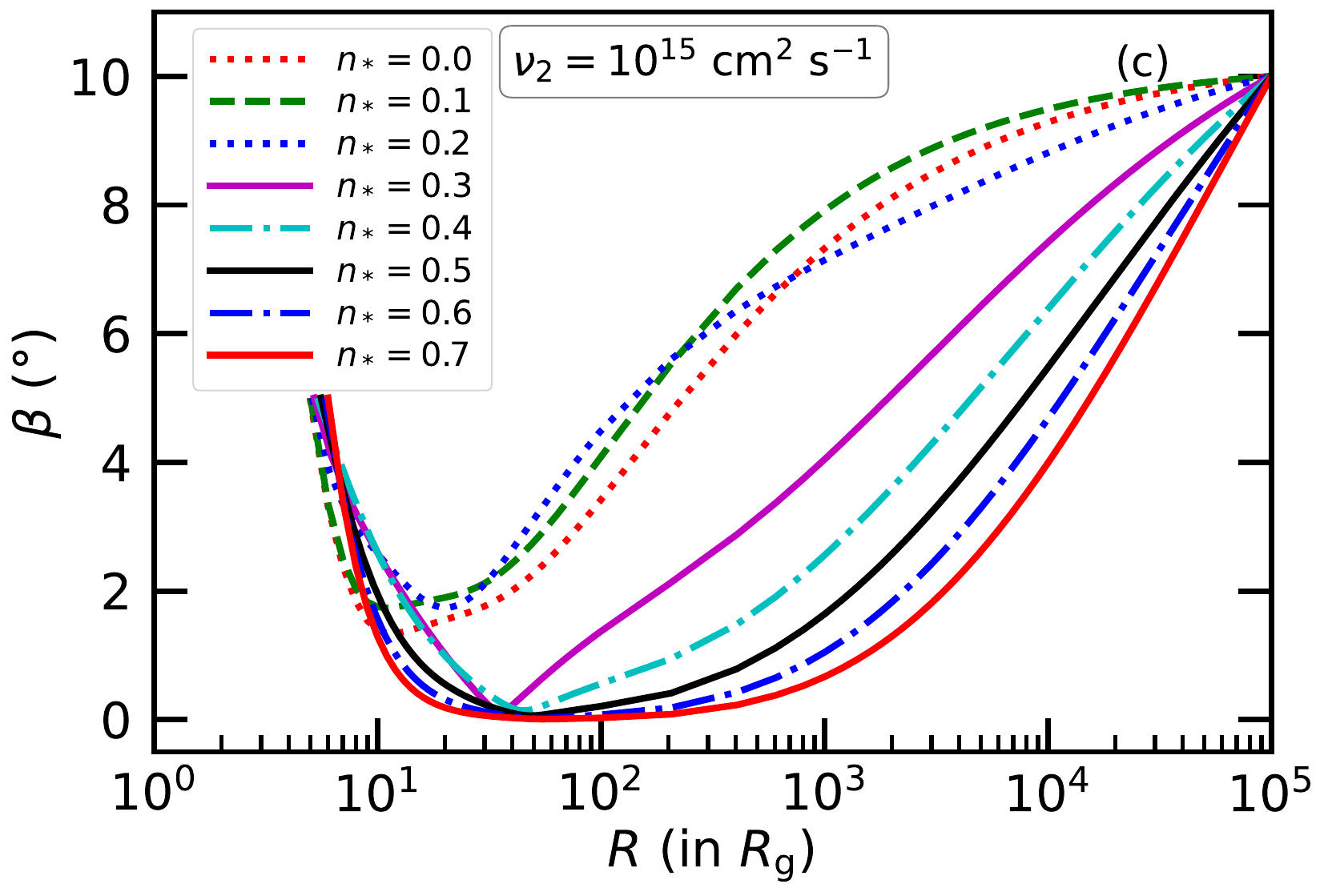}
	\includegraphics[width=0.49\textwidth]{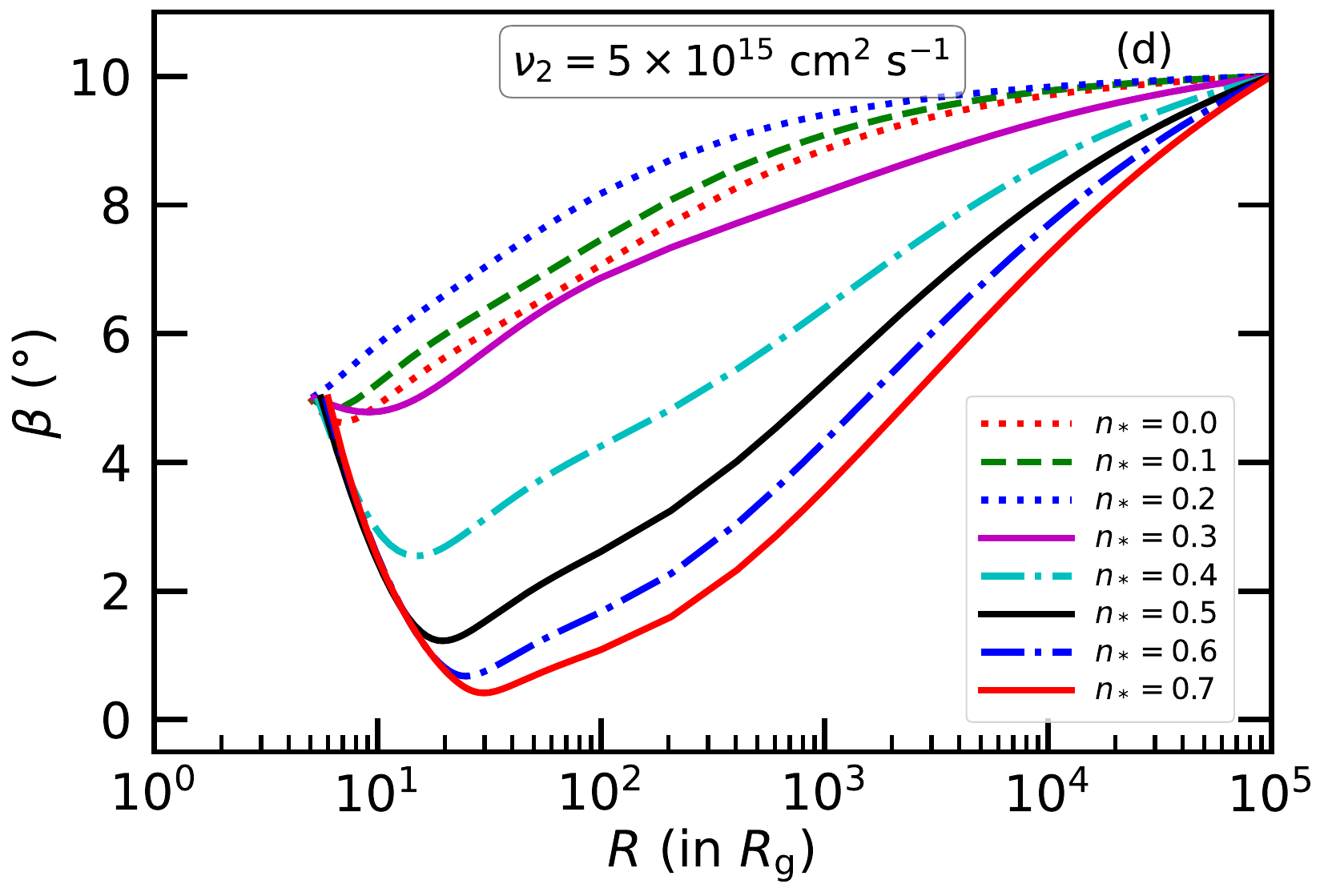}
	\caption{Variation of $\beta$ with $R$ for different values of $n_*$ with a fixed $a_*=0.3$, $M=10M_{\odot}$, $\eta=0.25$, $\beta_i=5^{\circ}$  and $z_{\rm in}=0.75$. The value of $\nu_2$ changes in the different panels as mentioned in the inset. See \S~ \ref{s4.2} for details.}
	\label{beta_r_nu_a0p3}
\end{figure}

Figure \ref{beta_r_nu_a0p3} depicts the variation of the tilt angle ($\beta$) with radial distance ($R$) for different values of $\nu_2$ but similar to Figure \ref{beta_r_nu_n0p3}.  
The various curves in each panel represent different values of $n_*$, ranging from $0.0$ to higher values with an interval of $\Delta n_* =0.1$ with a fixed value of $M = 10 M_{\odot}$ and $a_*=0.3$.
For $\nu_2=10^{14}$ cm$^2$ s$^{-1}$, the disk remains mostly aligned along the equatorial plane for any values of $n_*$. For $\nu_2=5\times10^{14}$ cm$^2$ s$^{-1}$, the disk tends to align along the equatorial plane for $n_* > 0.4$. For $\nu_2=10^{15}$ cm$^2$ s$^{-1}$, the disk aligns beyond $n_* > 0.5$, while for $\nu_2=5\times10^{15}$ cm$^2$ s$^{-1}$, the disk remains mostly tilted for any values of $n_*$. Here also, as the value $\nu_2$ increases, the viscous torque starts to dominate over the LT torque for a fixed set of $(a_*,n_*)$. In all the panels of figure \ref{beta_r_nu_a0p3}, the solid magenta curve shows the tilt profile for $a_*=n_*$. Remarkably, the curves for $n_* < a_*$ are located at the upper portion of the $xy$ plane, whereas the curves are located at the lower portion of the same plane for $n_* > a_*$, in the case of larger $R$. This indicates that the LT torque becomes stronger for the higher value of $n_*$ in that range of $R$. This does not only help to align the disk in the equatorial plane, but the alignment of the disk is also extended to the outer portion of the disk significantly. This could be clear by comparing the first and last terms of eq. (\ref{approx}), as it varies $\sim a_*/R^3$ and $\sim n_*^2/R^{5/2}$ respectively.
However, the above mentioned trend is not necessarily true at the lower $R$ near the inner boundary, as the dominant LT torque term switches from being the $a_*$ term to the $n_*$ term at $R_0$. Note that although the switching of LT torque occurs at $R_0$, the switching of the curves does not occur exactly at $R_0$. This is because 
$\Omega_{\rm p}$ is a function of $a_*$ and $n_*$. On the other hand, $\beta(R)$ is not only a function $a_*$ and $n_*$, but it is also largely dependent on $\nu_2$ and some other parameters, e.g., $M_{\rm BH}, ~\alpha, z_{\rm in}$, etc. Thus, the nature of $\Omega_{\rm p}$ and $\beta$ do not change in a similar manner always. Due to the same reason, the cross-over of $\beta$ curves does not occur exactly at $R_0$.
In the absence of $n_*$ (red dotted curve), the disk might be aligned in the equatorial plane for small viscosity (see panel (a)), but it is not aligned for the comparatively higher viscosity (see panels (b)-(d)). It is evident from panels (a)-(d) that there is a tendency of the alignment of disk in the equatorial plane for the higher values of $n_*$. It indicates that the higher values of $n_*$ play a significant role in the alignment, compared to $a_*$. It is because the last term ($\sim n_*^2/R^{5/2}$) of eq. (\ref{approx}) dominates over the first and second terms of the same equation. It is interesting to see the behavior of the dotted blue curves in panels (b) and (c) of figure \ref{beta_r_nu_a0p3}. As the difference between the value of $a_*$ (e.g., $a_*=0.3$) and $n_*$ (e.g., $n_*=0.2$) is small ($\sim 0.1$), the overall LT torque remains small for $R < 100 R_g$ and $R < 300R_g$ for panels (b) and (c), respectively. Note that $R_0$ occurs at $R_0 \sim 60 R_g$ for $(a_*, n_*)=(0.3, 0.2)$. Thus, the viscous torque dominates over the LT torque and the disk tends to be more tilted compared to the dashed green and dotted red curves around $R_0$. On the other hand, the LT torque due to $n_*$ starts to dominate over the LT torque due to $a_*$ for the larger $R$, and, hence, the overall torque dominates over the viscous torque. Thus, the disk tries to be aligned and, thereby, the dotted blue curves bent and cross the dashed green and dotted red curves at larger $R$ in panels (b) and (c). The dotted blue curve in panel (d) indicates that the viscous torque dominates over the overall LT torque in almost the whole range of $R$. Thus, one cannot see the tendency of disk alignment.

\begin{figure}[h]
	\begin{center}
		\includegraphics[width=0.49\columnwidth]{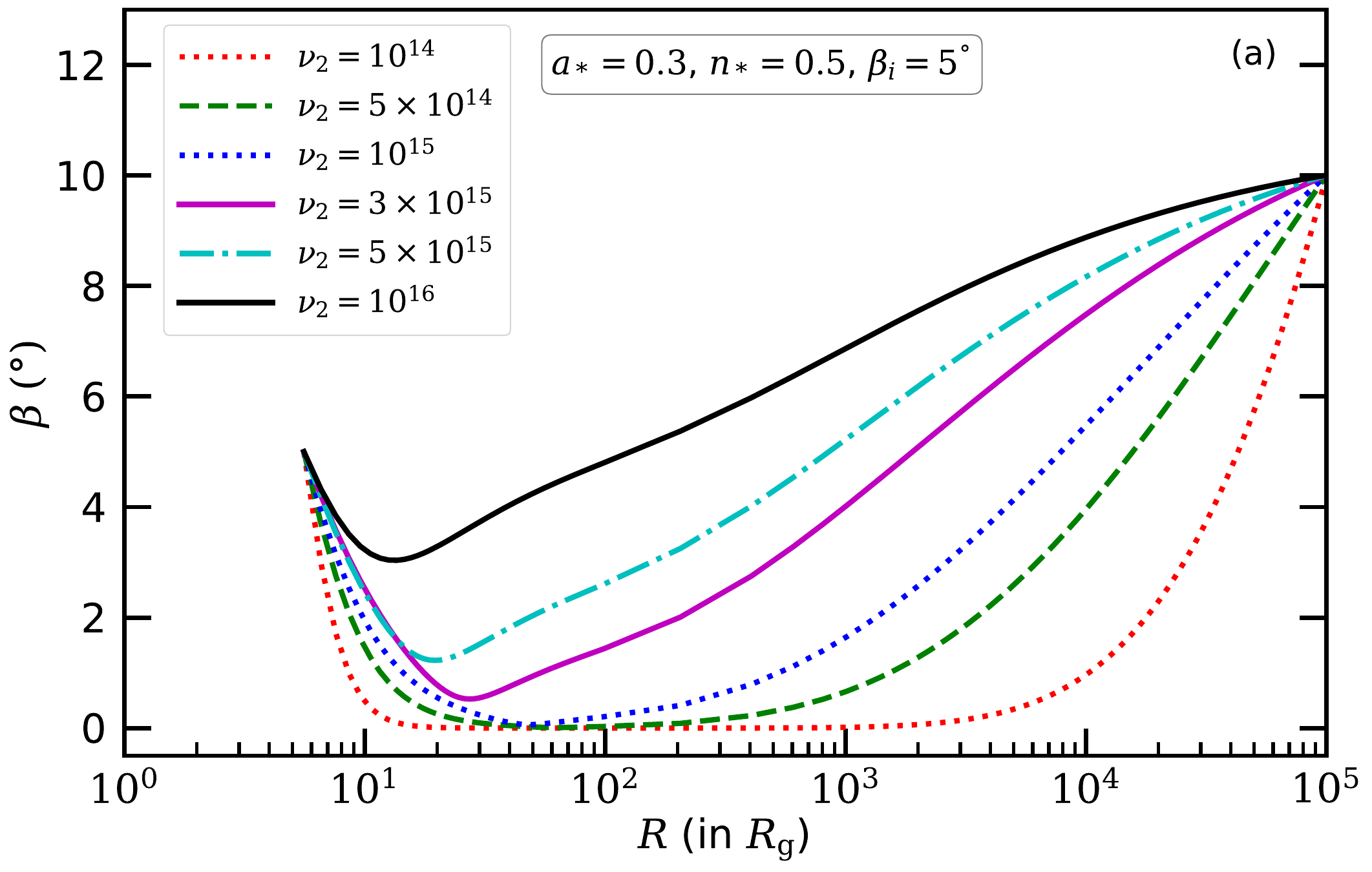}
		\includegraphics[width=0.49\columnwidth]{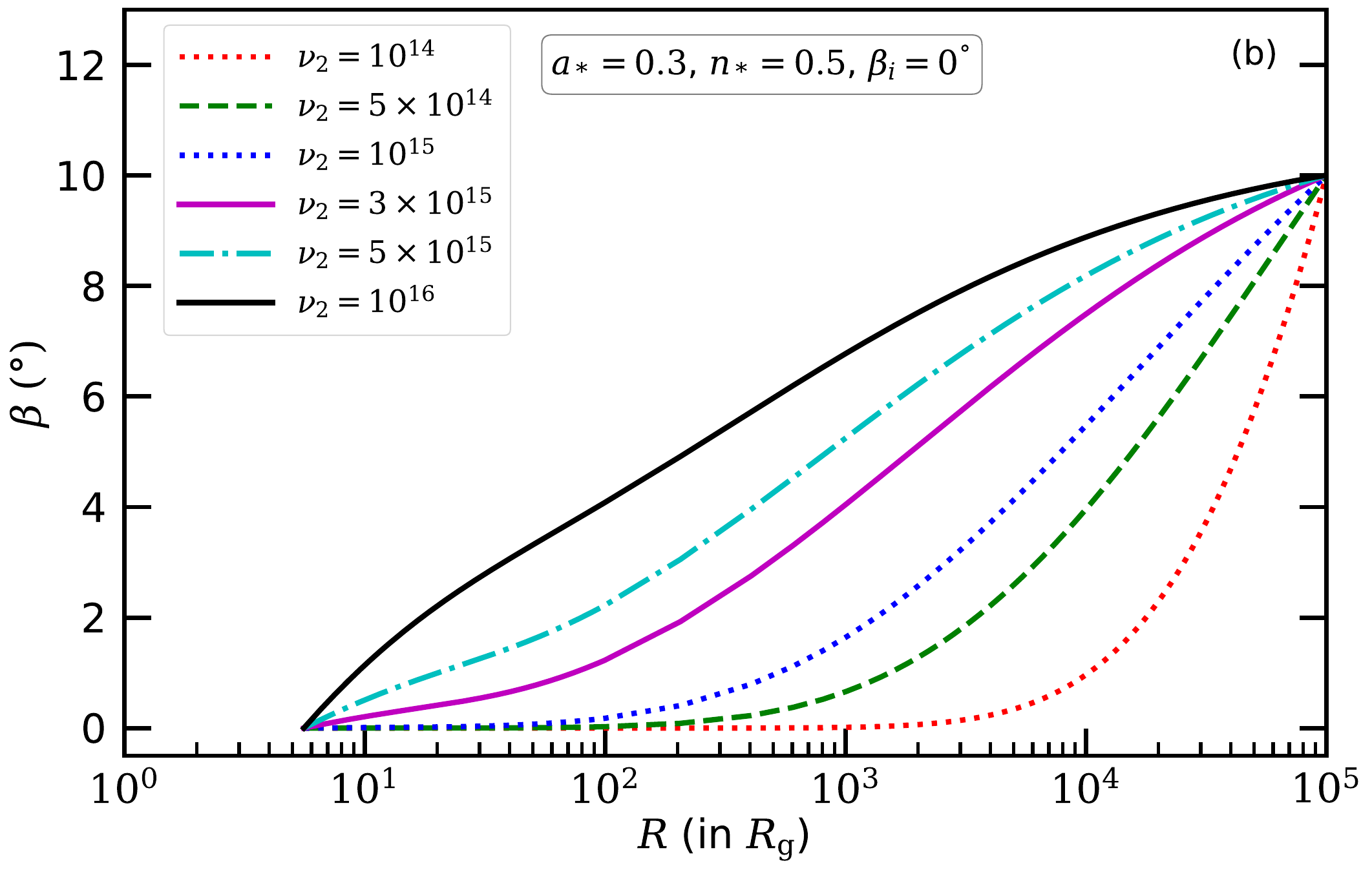}
		\includegraphics[width=0.49\columnwidth]{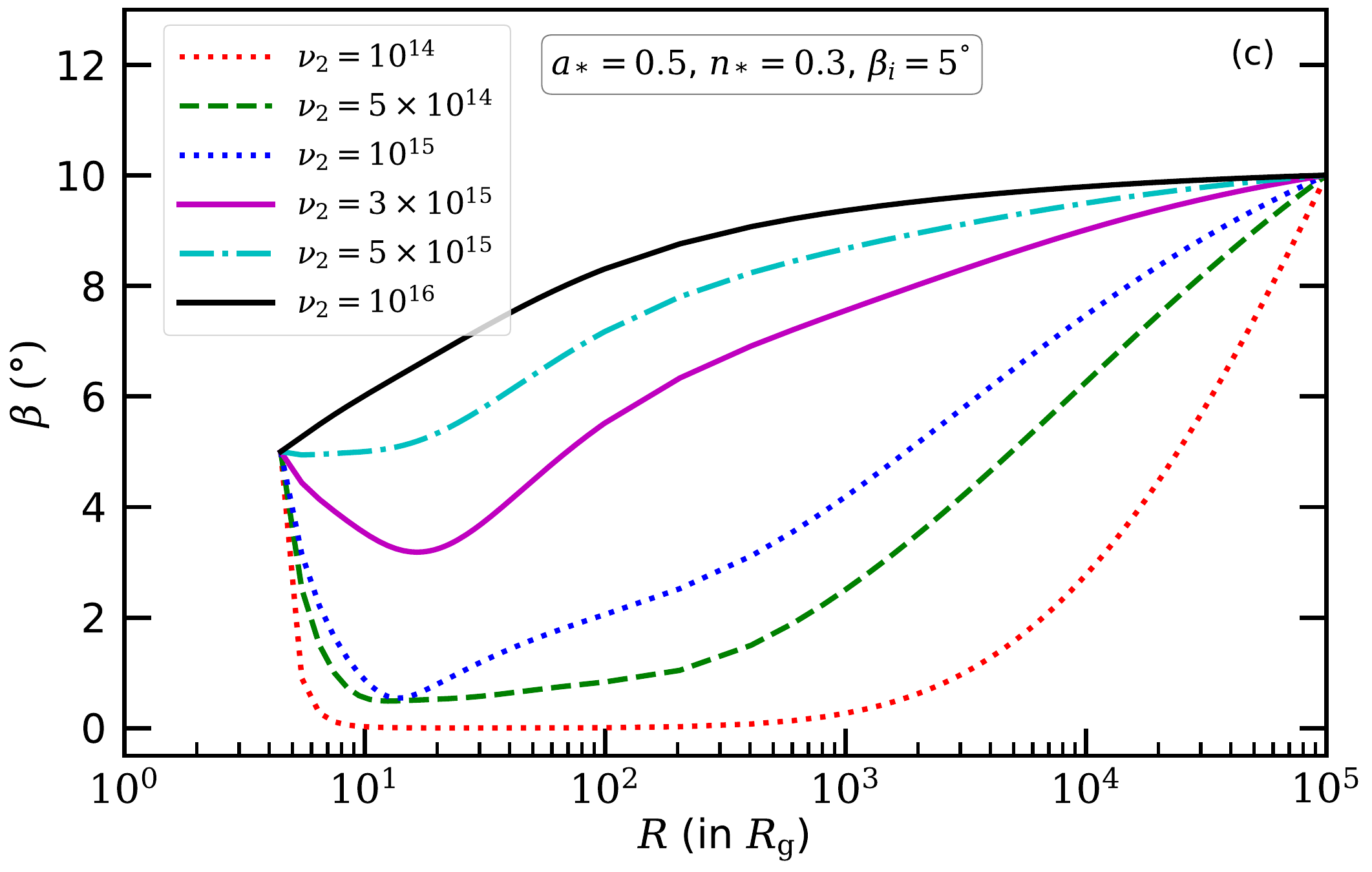}
		\includegraphics[width=0.49\columnwidth]{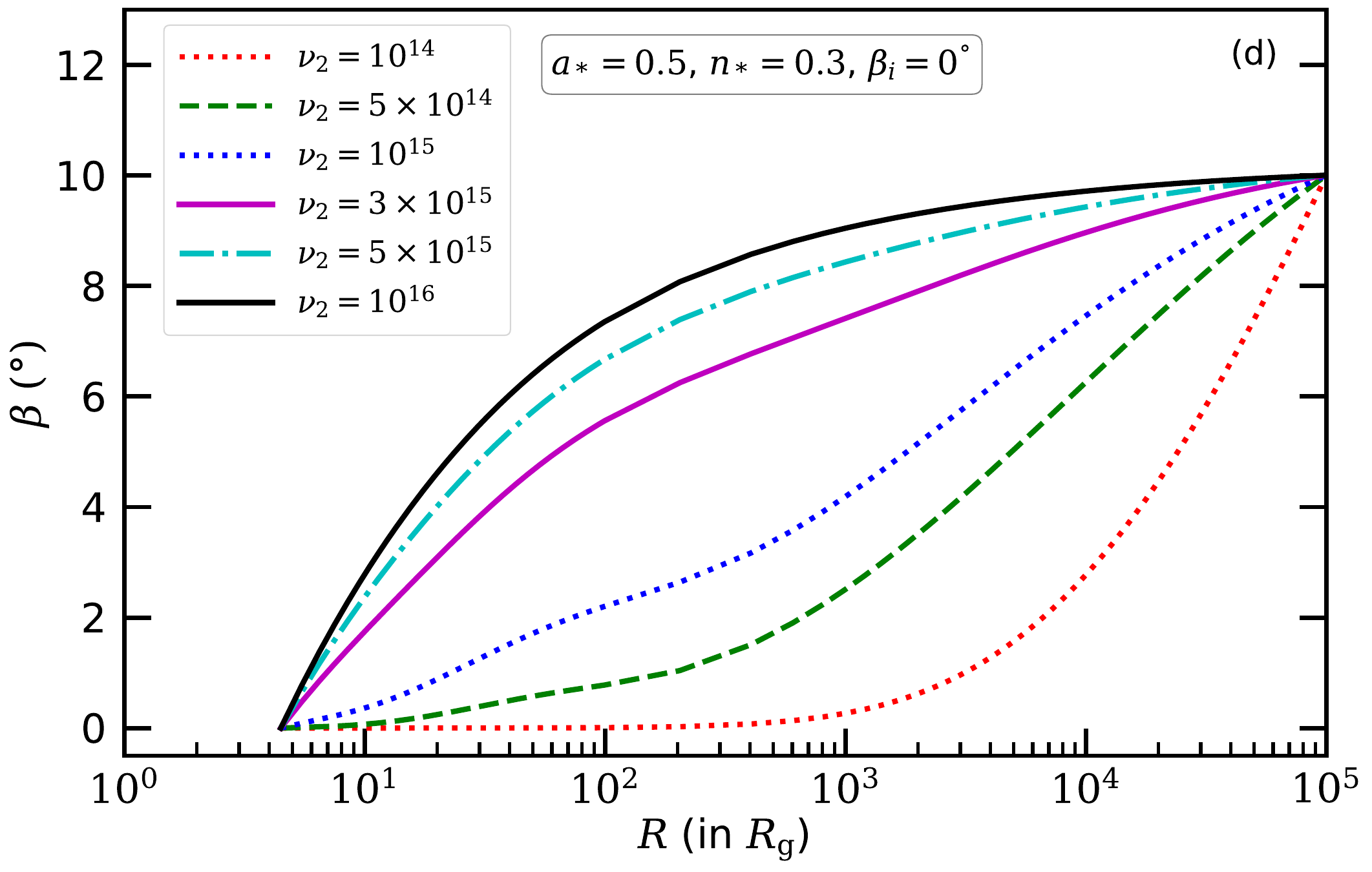}
		
	\end{center}
	\caption{Variation $\beta$ with $R$ for different values of $\nu_2$ with a fixed $M=10M_{\odot}, \eta=0.25$  and $z_{\rm in}=0.75$. The values of $a_*$ and $n_*$ are mentioned in the inset. See \S~ \ref{s4.2} for details.
	}
	\label{beta_r_nu2_a_n}
	
\end{figure}

Figure \ref{beta_r_nu2_a_n} portraits the variation of $\beta$ with respect to the radial coordinate for different values of $\nu_2$ with some fixed parameters: $M=10M_{\odot}, \eta=0.25$  and $z_{\rm in}=0.75$. The panels (a) and (b) depict scenarios where $(a_*,n_*)=(0.3,0.5)$ with $\beta_i=5^{\circ}$ and $0^{\circ}$, respectively. Similarly, the panels (c) and (d) represent cases with $(a_*,n_*) = (0.5,0.3)$ with $\beta_i$ values of $5^{\circ}$ and $0^{\circ}$, respectively. Studying all the panels, we can infer that, the inner disk is more tilted for large $\beta_i$. Moreover, it is also seen that the tilting of the disk increases with the increment of $\nu_2$.

\begin{figure}[h!]
	\includegraphics[width=0.49\textwidth]{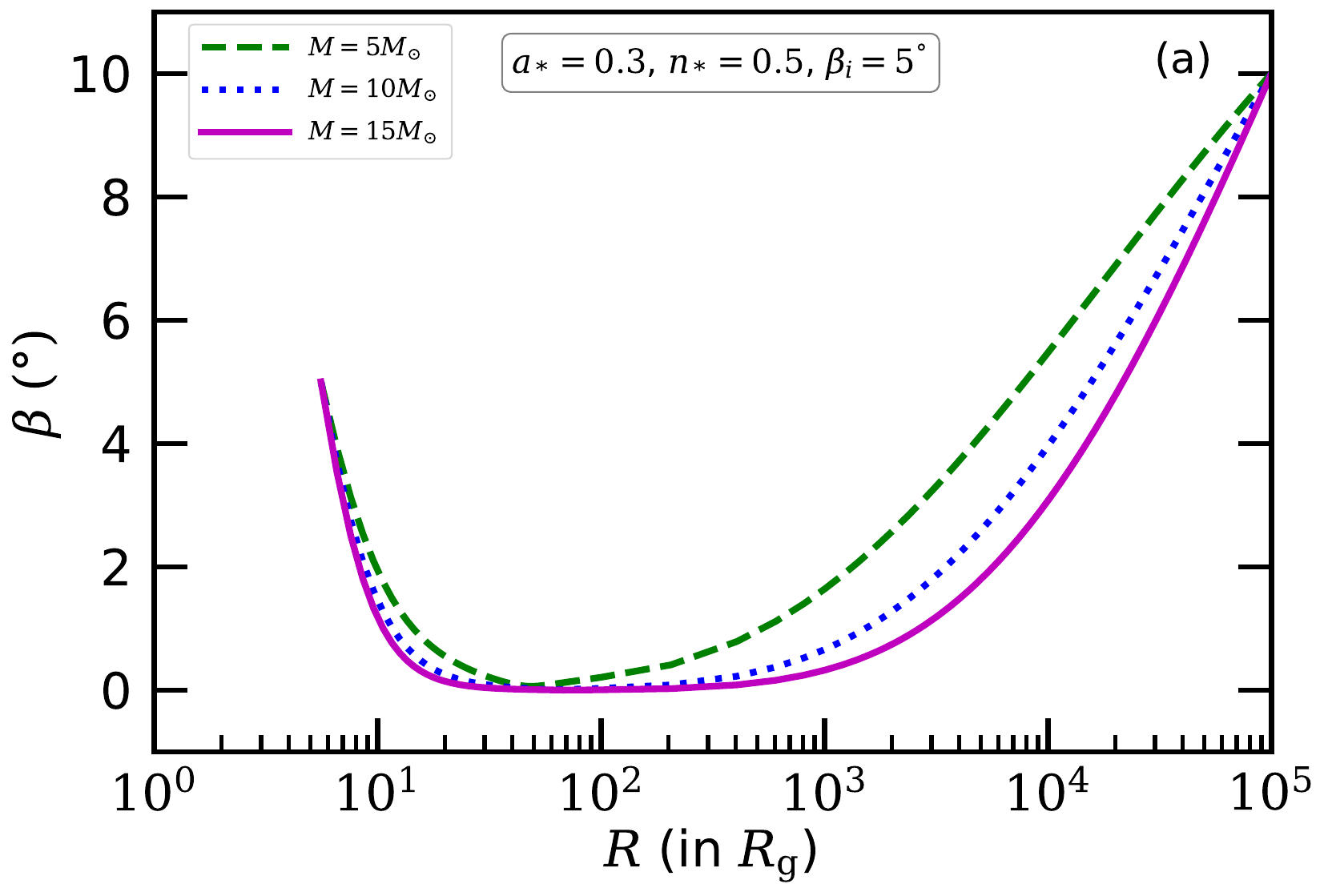}
	\includegraphics[width=0.49\textwidth]{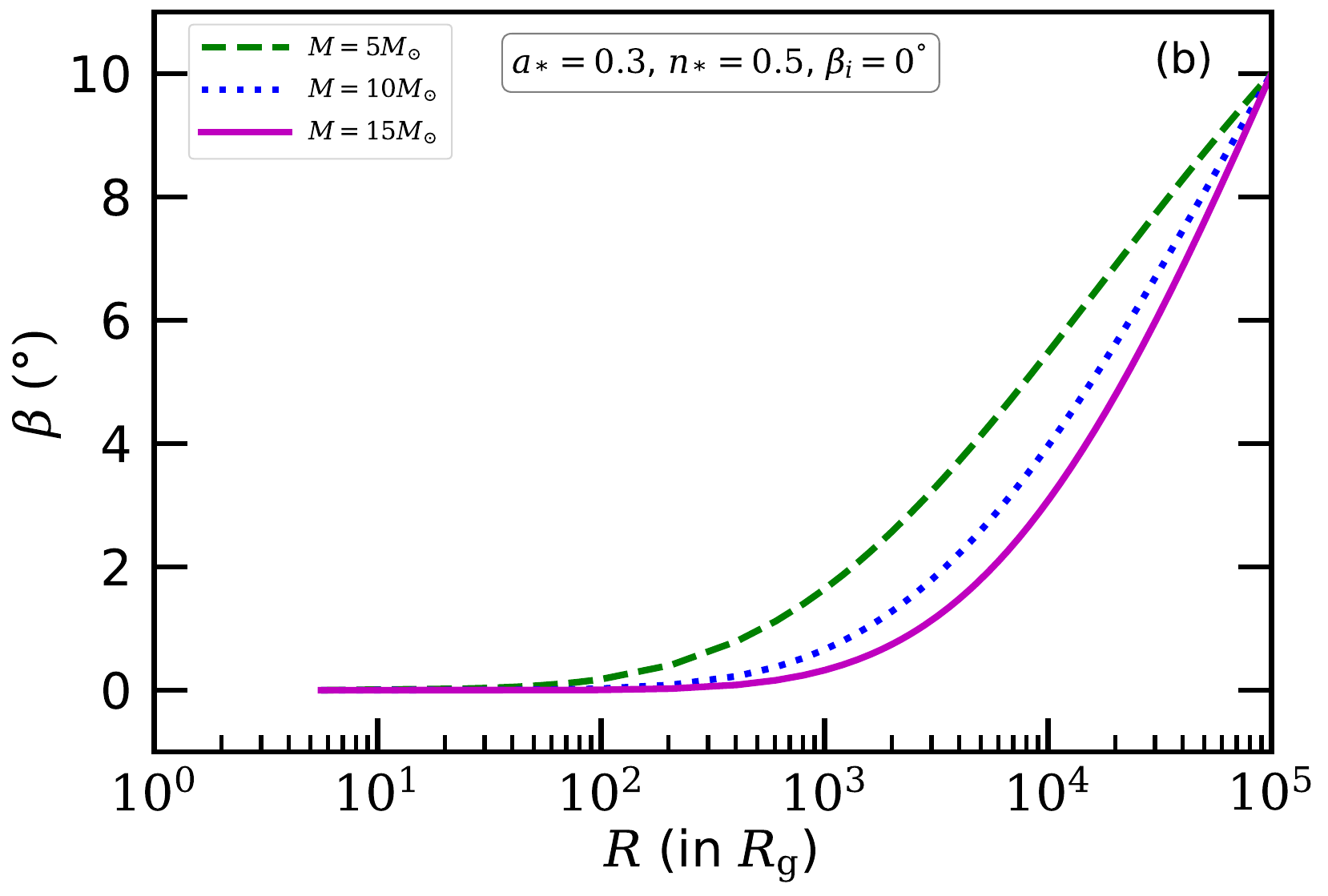}
	\includegraphics[width=0.49\textwidth]{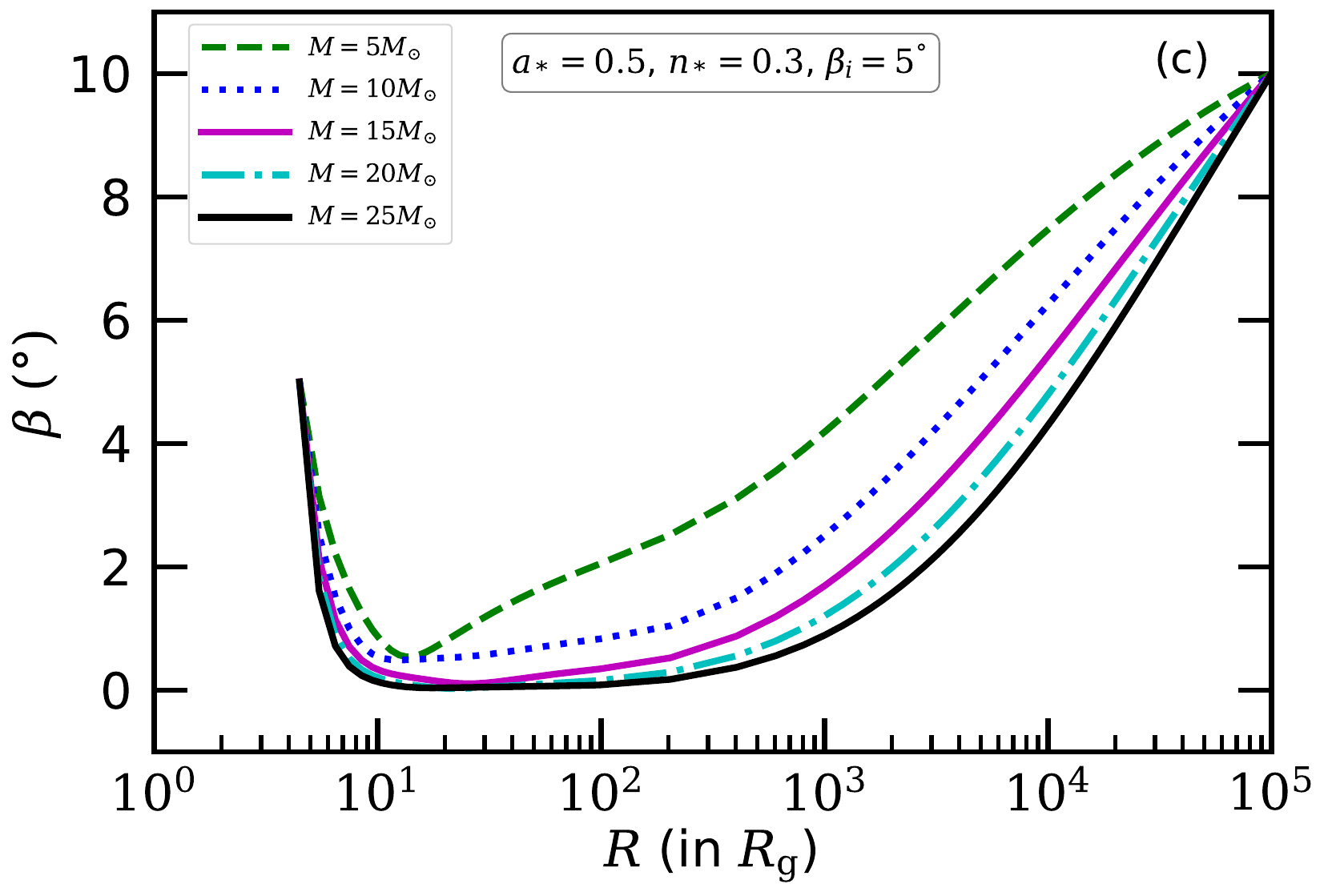}
	\includegraphics[width=0.49\textwidth]{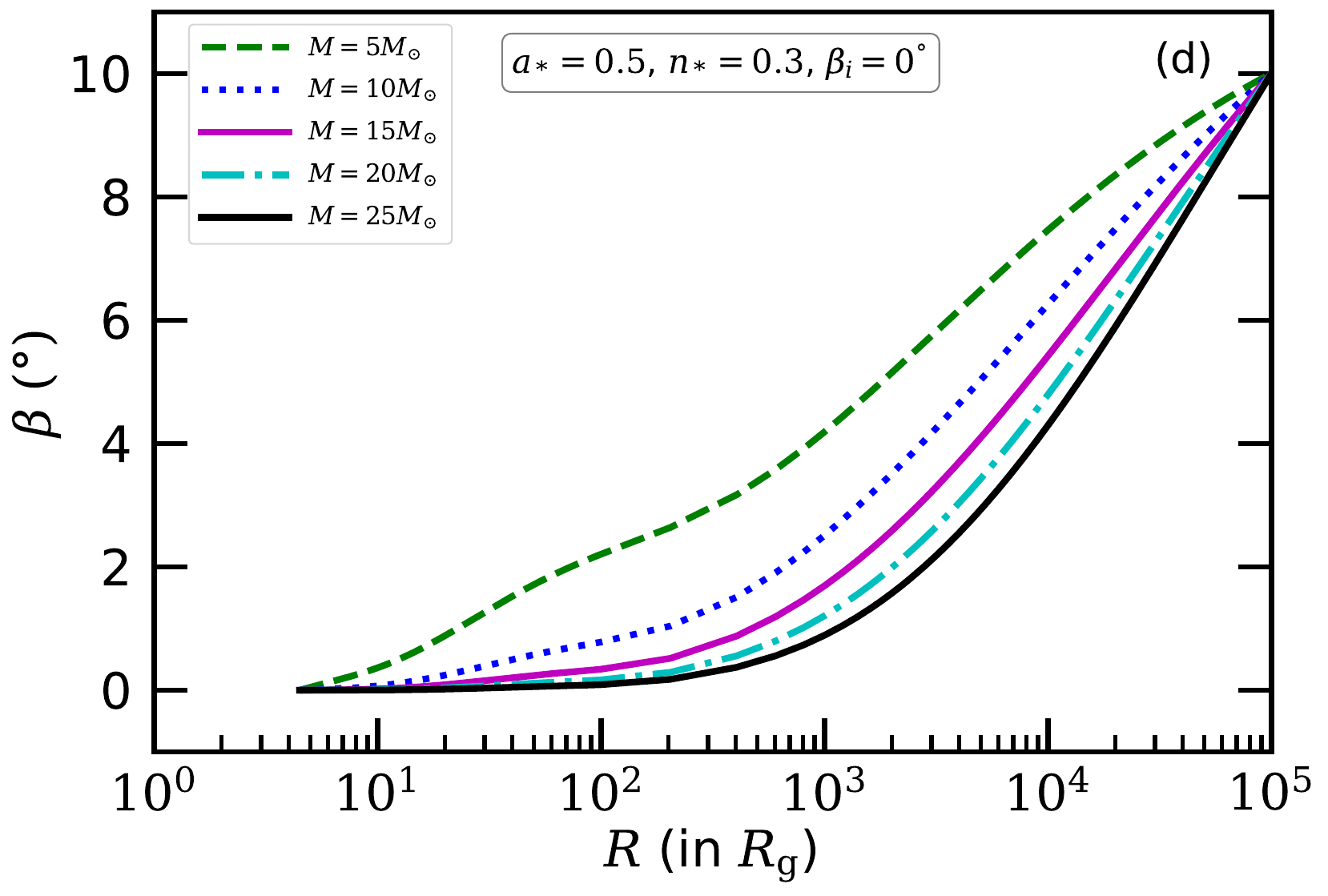}
	
	\caption{Variation of $\beta$ with $R$ for different values of black hole mass ($M$) for  a fixed $\nu_2=5 \times 10^{14}$ cm$^2$s$^{-1}$, $\eta=0.25$  and $z_{\rm in}=0.75$. The values of $a_*$ and $n_*$ are mentioned in the inset. See \S~ \ref{s4.2} for details.}
	\label{Beta_r_m_5nu214}
\end{figure}

Figure \ref{Beta_r_m_5nu214} showcases the variation of $\beta$ with respect to the radial coordinate for different values of $M$. Here we fix $\nu_2=5\times10^{14}$ cm$^2$ s$^{-1}, \eta=0.25$  and $z_{\rm in}=0.75$. The panels (a) and (b) correspond to scenarios where $(a_*,n_*)=(0.3, 0.5)$ with $\beta_i=5^{\circ}$ and $0^{\circ}$, respectively. Similarly, the panels (c) and (d) represent cases with $(a_*,n_*)=(0.5,0.3)$, with $\beta_i=5^{\circ}$ and $0^{\circ}$, respectively. Across all the subfigures, it is observed that the LT torque dominates for higher black hole masses. This observation highlights the significance of the black hole mass in determining the dominance of the LT torque and its effect on aligning the accretion disk with the black hole's spin axis.

\begin{figure}[h!]
	\includegraphics[width=0.49\textwidth]{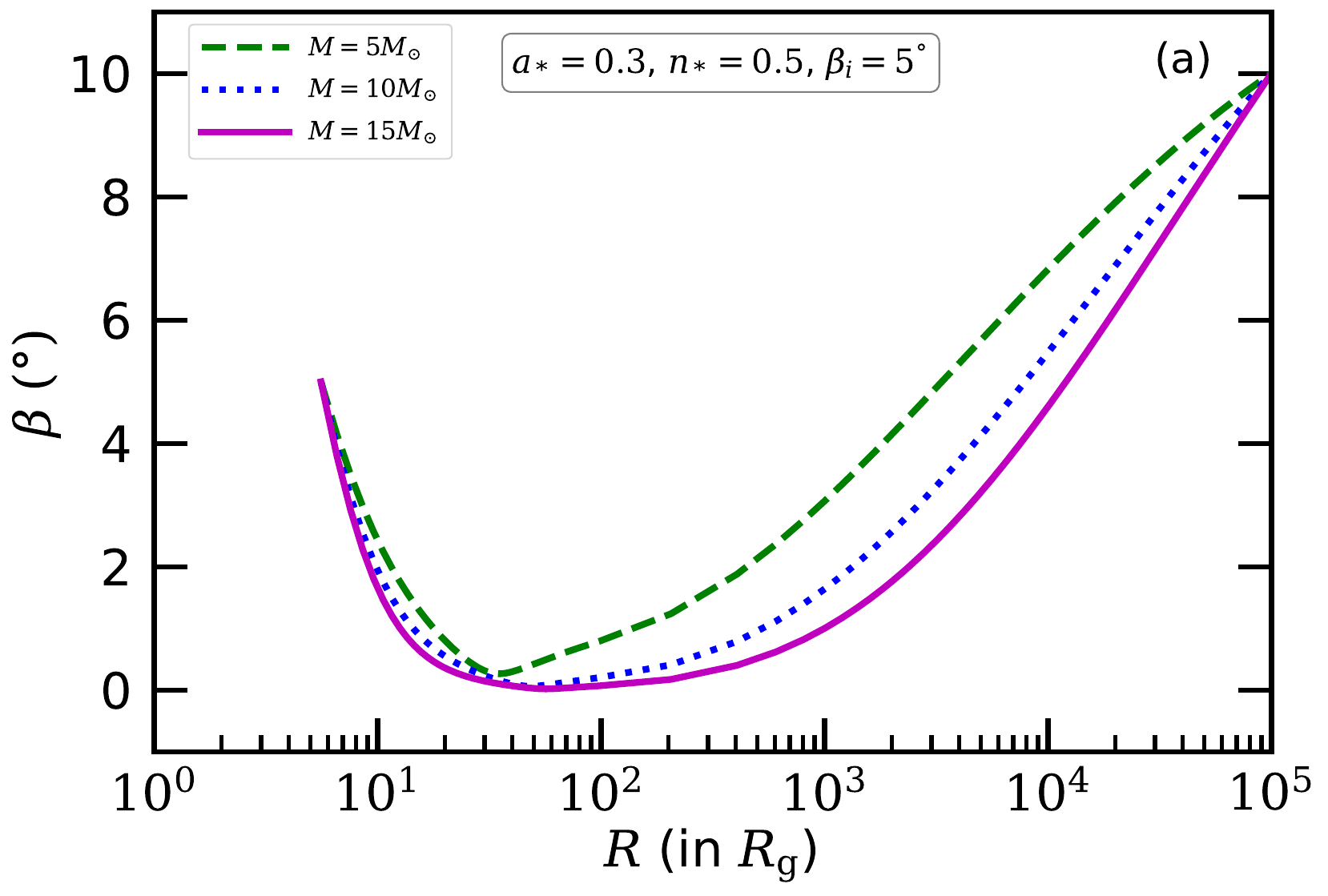}
	\includegraphics[width=0.49\textwidth]{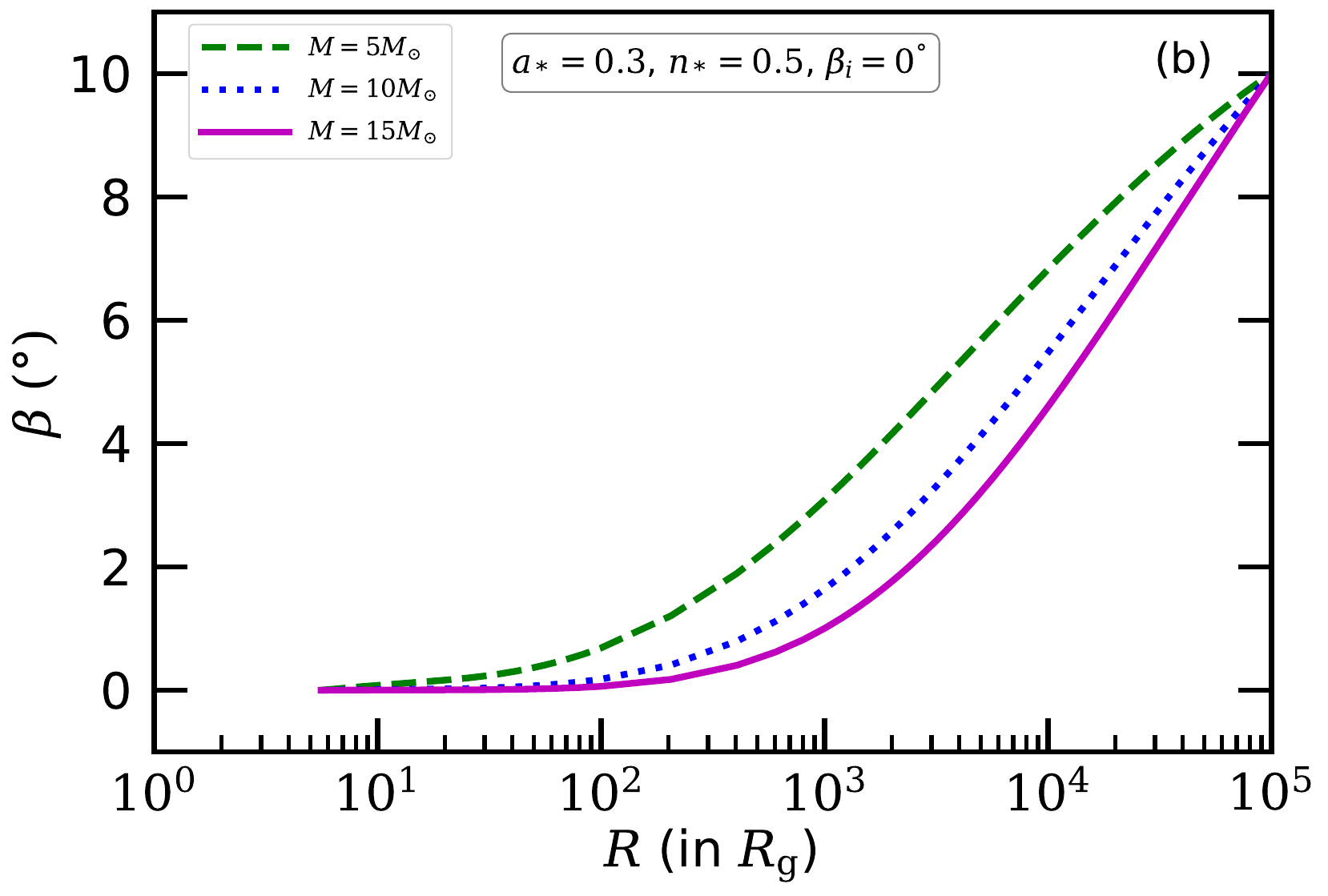}
	\includegraphics[width=0.49\textwidth]{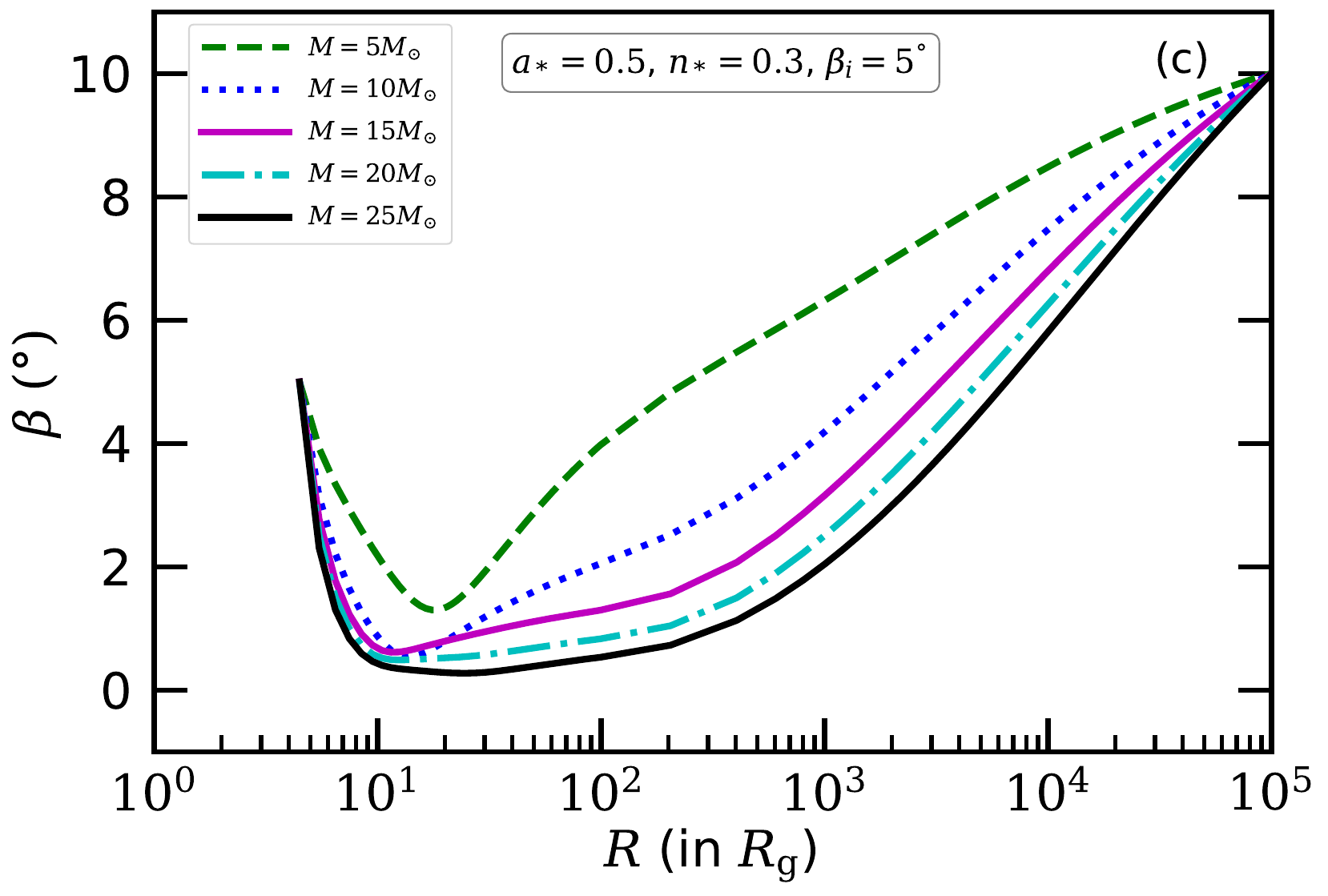}
	\includegraphics[width=0.49\textwidth]{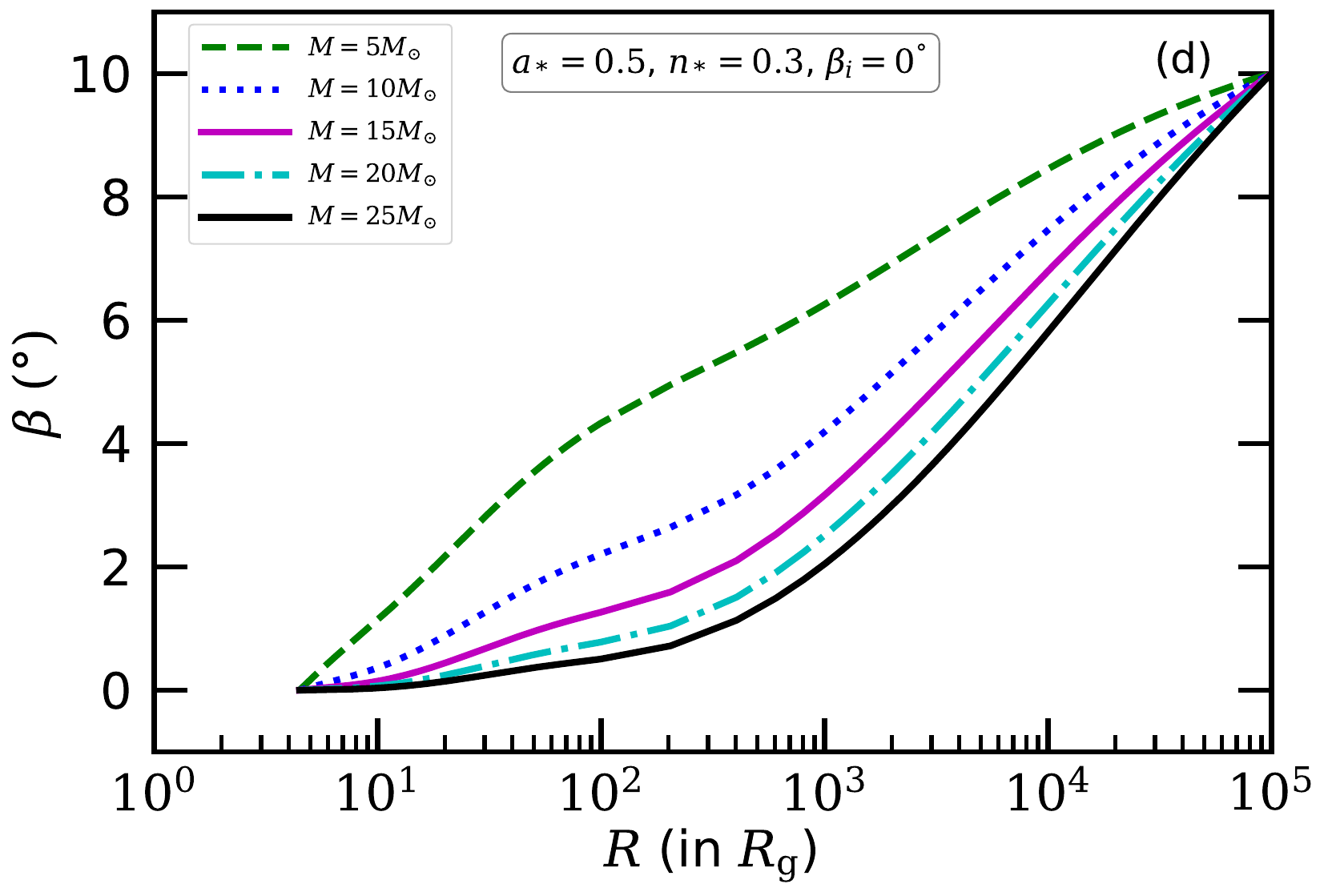}
	
	\caption{Similar to figure \ref{Beta_r_m_5nu214} but for a fixed $\nu_2=10^{15}$ cm$^2$ s$^{-1}$. See \S~ \ref{s4.2} for details.}
	\label{Beta_r_m_nu215}
\end{figure}

Figure \ref{Beta_r_m_nu215} is similar to figure \ref{Beta_r_m_5nu214} but  for a different value of $\nu_2$, i.e., $\nu_2=10^{15}$ cm$^2$ s$^{-1}$. Comparing figure \ref{Beta_r_m_5nu214} and figure \ref{Beta_r_m_nu215}, we can infer that when the viscosity parameter $\nu_2$ (representing viscous torque) is higher, a more massive black hole (higher $M$) is needed to align the disk. This observation highlights the complex relation among the black hole mass, viscous torque, and the alignment dynamics of the accretion disk.

\subsection{\label{s4.3}Warp radius of the disk}
It is seen from the above discussion that a disk is partially aligned for a set of parameter values. In such a case, one can consider a radius $R_{\rm align}$ up to which the disk remains aligned. It is higher for the higher values of the LT effect. In our computation, we consider $R_{\rm align}$ as the radius up to which the tilt angle is less than $0.01^{\circ}$ following \cite{Banerjee-etal2019}. The characteristic warp radius ($R_{\rm warp}$) is defined as the radius inside which the LT effect dominates \cite{Banerjee-etal2019}. It is calculated by equating the timescale for warp diffusion (i.e., $R^2/\nu_2$) and the local LT precession timescale (i.e., $1/|\O_{\rm p}|$). Thus, one can solve the following equation
\begin{eqnarray}
\Bigg|\left(\frac{2a_*R_g}{R}-\frac{3a_*^2 R_g^{3/2}}{2R^{3/2}}-\frac{2n_*^2R_g^{1/2}}{R^{1/2}}(1-2R_g/R)^2\right)\Bigg|=\f{\nu_2}{cR_g},
 \label{warp}
\end{eqnarray}
for $R=R_{\rm warp}$ to obtain the warp radius for our case. It gives $R_{\rm warp}=2a_*cR_g^2/\nu_2$ for a Kerr black hole, which resembles eq. (45) of \cite{Banerjee-etal2019}. When the LT effect due to $n_*$ (3rd term of eq. \ref{warp}) dominates over the LT effect due to $a_*$ (1st term of eq. \ref{warp}), one has to solve it by considering the negative sign in the right hand side of eq. (\ref{warp}). Otherwise, it has to be considered as positive. 
Let us take an example from the plots of the panel (d) of figure \ref{beta_m_nu2} which corresponds to $\beta (R)$ for $M=10M_{\odot},~ a_*=0.4$ and $\nu_2=10^{15}$ cm$^2$ s$^{-1}$ with four different values of $n_*$. For $(a_*,n_*)=(0.4,0.0)$, we obtain $R_{\rm warp} \sim 34 R_g$ by considering the above-mentioned positive sign. For this set of parameters, the disk cannot be in the situation where tilt angle $\beta < 0.01^{\circ}$. This implies that $R_{\rm align}$ does not arise in this case. Therefore, we can say that the LT torque cannot overrule the viscous torque to make the disk aligned in the equatorial plane. Similarly, for $(a_*,n_*)=(0.4,0.4)$ and $(0.4,0.5)$, we obtain $R_{\rm warp} \sim 110 R_g$ and $\sim 411 R_g$ respectively, by considering the negative sign in eq. (\ref{warp}). For $(a_*,n_*)=(0.4,0.4)$ and $(0.4,0.5)$, we obtain $R_{\rm align} \sim 53 R_g$ and $\sim 52 R_g$ respectively. The values of $R_{0}$ are obtained as $\sim 12 R_g$ and $\sim 8 R_g$, respectively, for the above mentioned cases. One obtains $R_{\rm warp} \sim 13 R_g$ for $(a_*,n_*)=(0.4,0.3)$ by considering the positive sign. Note that the LT precession vanishes at $R_0 \sim 24.5 R_g$ in this particular case as mentioned earlier. It indicates that the LT effect lost its control over the disk for $R > 13 R_g$. Let us consider another sample from panel (a) of figure \ref{beta_r_nu2_a_n}. In this case, $R_0$ is located at $\sim 6.8 R_g$ for $(a_*,n_*)=(0.3,0.5)$. However, depending on the numerical values of $\nu_2$, the locations of $R_{\rm warp}$ and $R_{\rm align}$ would be different. For instance, if we take $\nu_2=10^{14}$ cm$^2$ s$^{-1}$, we obtain $R_{\rm warp}\sim 48890 R_g$ and $R_{\rm align}\sim 1127 R_g$. On the other hand, $R_{\rm warp}\sim 1861 R_g$ and $R_{\rm align}\sim 73 R_g$ for $\nu_2=5\times10^{14}$ cm$^2$ s$^{-1}$. Similarly, for $\nu_2=10^{15}$ cm$^2$ s$^{-1}$, we obtain $R_{\rm warp}\sim 431 R_g$ and $R_{\rm align}\sim 49 R_g$. If we further increase the viscosity to $\nu_2=3\times10^{15}$ cm$^2$ s$^{-1}$, $R_{\rm warp}$ and $R_{\rm align}$ cannot be obtained, as $R_{\rm warp} < R_{\rm ISCO}$. Thus, we see that if we increase the viscosity, the viscous torque leads over the LT torque after a certain value of $\nu_2$. As a result, the $R_{\rm warp}$ and $R_{\rm align}$ shift in the lower $R$, and beyond a particular $\nu_2$, both of them cease to exist. This is also clear from the curves of the panel (a) of figure \ref{beta_r_nu2_a_n}. Moreover, due to the higher viscous torque compared to the overall LT torque, it cannot take over the control of the disk for $R > R_0$. Eventually, the viscous torque dominates the overall LT torque, and the $\beta(R)$ curve starts to move upward.
The relation between $R_{\rm align}$ and $R_{\rm warp}$ was found to be $R_{\rm align}=0.165R_{\rm warp}$ \cite{nat} for the Kerr black hole. The proportionality factor was later obtained as $\sim 0.094$ in \cite{Banerjee-etal2019}. However, 
due to the non-monotonous behaviour of the $\beta (R)$ curves in the KTN black hole, the proportionality factor between $R_{\rm align}$ and $R_{\rm warp}$ does not remain fixed across the parameter space explored in this paper. The proportionality factor roughly varies within $0.03 - 0.45$, as obtained from our computations.

\subsection{\label{s4.4}Justification for using the approximated expression of LT precession to calculate $\beta(R)$}~

We have considered here the approximate LT precession expression for the KTN black hole and, hence, considered the numerical values of $a_*$ and $n_*$ up to $0.7$ to calculate the tilt profiles. In general, the approximate expression ($\Omega_{\rm LT} \sim 2a_*M^2/R^3$) of LT precession in Kerr black hole is used upto $a_* \rightarrow 1$ in many papers \cite{cap, fra1, fra2, mar, nel, nat, li1, li2, Banerjee-etal2019, bcb19}. This is because the above-mentioned expression is not only applicable for the slowly-spinning ($a_* << 1$) Kerr black hole, but it is also applicable to the Kerr black holes with high spin, if $a_*$ is small enough than $R/M$ (i.e., $a/R$ is much less than $1$). This is also clear from eq. (\ref{ltk}). After a careful inspection, we find that if one uses that expression ($\Omega_{\rm LT} \sim 2a_*M^2/R^3$) for the Kerr black hole with $a_* < 1$, no qualitative differences could be found in the tilt profiles. However, a quantitative difference (maximum upto $\sim 15\%$) in the tilt angle could appear at $R \sim 5R_g$ for $a_* = 0.7$, and tends to zero for $R > 9R_g$. For $a_* < 0.7$, the above-mentioned quantitative difference decreases significantly from $\sim 15\%$.
It is expected because the LT effect in a Kerr black hole is mainly governed by the inverse cube law of distance (i.e, $a/R^3$) for $a_* << R/M$, as explained above. Now, since the value of $a_*$ for a Kerr black hole is comparatively much less than $R/M$ (the minimum value of $R$ is $R_{\rm ISCO}$) in reality, the difference in the numerical values obtained using the exact expression and the approximate expression of LT precession does not differ significantly. 
That is why, the approximate expression of LT precession is applied \cite{cap, fra1, fra2, mar, nel, nat, li1, li2, Banerjee-etal2019, bcb19} even to the higher $a_*$ values for calculating the tilt profile.  Note that the qualitative and substantial quantitative differences can be seen only for those $a_*$ values which represent the Kerr naked singularities, i.e., $a_* > 1$. This is because, $a_*$ becomes comparable to $R/M$, or, even $a_* > R/M$ at $R_{\rm ISCO}$ and close to that radii for some values of $a_*$ (see Sec. VII of \cite{ckp}). Due to the same (above-mentioned) reason, we have also used here the approximate expression of LT precession (eq. \ref{approx}) for calculating the tilt profile and applied it to the KTN black hole with the value of $n_*$ up to $0.7$ along with $a_* = 0.7$. In addition, eq. (\ref{approx}) reveals that the dominant contribution (due to the presence of GMM) comes from $n^2/R^{5/2}$ by neglecting the higher order terms for $n_* << R/M$. 
Although no qualitative differences appear in the tilt profiles calculated using the exact and approximate expressions (i.e., for $a_* << R/M$ and $n_* << R/M$)  of LT precession frequency in the KTN black hole, a quantitative difference (maximum upto $\sim 11\%$) in the tilt angle could appear at $R \sim 8.4R_g$ for $(a_*,n_*) =(0.7, 0.7)$, and tends to zero for $R > 16R_g$. Comparatively for the lower values of $(a_*, n_*)$, the above-mentioned quantitative differences are expected to be less than $11\%$.
However, the qualitative and substantial quantitative differences are supposed to be seen if one of the parameters ($a_*, n_*$) or both are greater than or very close to $1$. In such a case, one has to use the full LT precession frequency expression (eq. \ref{omega_LT1}) to solve the coupled differential equations similar to eqs. (\ref{lxf}-\ref{lyf}), the left hand sides of which would also be obtained as a function of $a_*$ and $n_*$. 
This means that the left hand side of eq. (\ref{angular_steady_state}) should be modified. Actually, the assumptions that lead to eq. (\ref{angular_steady_state}) are valid only within the parameter space where the effects of general relativity are sufficiently weak. This enables us to express the angular frequency using the Newtonian formula. We demand $R > R_g$ in this instance, such that one can expand eq. (\ref{omega_LT1}) to obtain the expression of $\Omega_{\rm p}$ (eq. \ref{approx}), since $a_* / (R/R_g)$ is inevitably small in this regime.
The exact formulation of a tilted thin accretion disk around the Kerr and KTN spacetime (valid for both black holes and naked singularities \cite{ckp, Chakraborty-Bhattacharyya2018, cbgm2}) is necessary to study the tilted accretion disk around GRO J1655-40 and some other astrophysical collapsed objects, as discussed in section \ref{intro}. This is because it was shown in \cite{Chakraborty-Bhattacharyya2018} that the suitable parameter space for GRO J1655-40 is ($a_* > 1,  n_* > 1$), which has not been considered in this paper. This is in preparation and will be reported in the future elsewhere.

\section{\label{s5}Summary and Conclusion}

In this paper, we numerically solve the warped accretion disk equations in the viscous regime for a KTN black hole. Taking into account the inner disk contribution, we obtain the radial profile of the tilted disk around the said black hole starting from the outer edge of the disk upto ISCO radius. We analyze the radial profile of the tilted disk as a function of the several parameters from our numerical results. We find that the inner disk could be entirely misaligned for a reasonable range of parameter values, which could confront with the astrophysical observations. For specific values of $M$ and $\nu_2$, there are combinations of $(a_*, n_*)$, where the disk begins to align with the direction of the black hole spin, indicating that the Bardeen-Petterson effects could be observed. We summarise the key takeaways from the parameter sweep below.

\begin{itemize}

    \item The alignment of the disk strongly depends on the tug-of-war between the viscous torque (controlled by $\nu_2$) and the LT torque (controlled by $M$, $a_*$ and $n_*$). In case of the KTN black hole, the LT precession does not monotonically increase with the decrement of $R$ towards $R_{\rm ISCO}$, unlike a Kerr black hole. Instead, if the value of $R$ decreases from the outer orbit to ISCO, the modulus of LT precession frequency first increases, attains a peak, then decreases to zero, and increases again depending on the numerical values of $(a_*, n_*)$ and the location of ISCO. Therefore, it is evident that the interplay between the LT effects due to $a_*$ and $n_*$ is also important in this case.

    \item Since, the radial profile of $\beta$ is affected by the LT effect, the behavior of all the curves for $\beta (R)$ is non-monotonous. For the same reason, the proportionality factor between $R_{\rm align}$ and $R_{\rm warp}$ for the various $\beta (R)$ curves does not remain fixed for the KTN black holes. 
    
    \item It is interesting to see that the LT torue due to GMM is extended comparatively to the outer disk than the LT effect produced by the Kerr parameter, as it varies $R^{-5/2}$ in case of the KTN black holes.  

    \item The $\beta (R)$ curves for $n_* < a_*$ are located at the upper portion of the $n_* = a_*$ curve in the $\beta - R$ plane, whereas they are located at the lower portion of the same plane for $n_* > a_*$, in the case of larger $R$. This indicates that the LT torque becomes stronger for the higher value of $n_*$ in that range of $R$. However, the above-mentioned trend is not necessarily true at the inner disk, as the dominant LT torque term switches from being the $a_*$ term to the $n_*$ term at $R_0$.
    
    \item For every combination of $a_*$, $n_*$  and $M$ ($\nu_2$), there is a specific value of $\nu_2$ ($M$), above (below) which the inner edge of the disk remains misaligned.

\end{itemize}

Certainly, the study of inner accretion disk plays an important role to probe the strong gravity regime. Tilting of the inner accretion disk with respect to the black hole spin axis affects the spectral and timing properties of the X-ray emission through the LT precession, and therefore, it is useful to study the same in strong gravity regime. In fact, C-type low frequency quasi-periodic-oscillation (QPO) frequency is identified as the LT precision frequency \cite{sv98, sv99, inm}, and shown to be emerged from the inner accretion disk. Note that a tilted inner disk has been inferred from X-ray spectral and timing features of the accreting black hole H1743-322 \cite{in}. GRO J1655-40 is still misaligned \cite{martin}. These misalignment's could be explained employing the present formulation. More importantly, since the existence of GMM is hinted in GRO J1655-40 \cite{Chakraborty-Bhattacharyya2018} and M87* \cite{gcyl}, and even in the galaxy \cite{gov, rug}, our solution for the radial profile of the tilted disk around a KTN black hole could be useful to probe the strong gravity regime as well as the existence of GMM in nature. Although some degeneracies could appear in the tilt profiles between a Kerr black hole and a KTN black hole with some specific combinations of $(a_*, n_*)$ in some values of $R$, it is unlikely to match both of the tilt profiles exactly as a whole at each and every value of $R$ for these two different type of black holes. This is because the functional dependence of the LT precession frequency as well as the tilt profile on $a_*, n_*$ and $\nu_2$ for all values of $R$ could not be exactly the same for a Kerr and a KTN black hole. However, probing the strong gravity regime with the accurate measurements of the tilt profile can help to break the degeneracies (if any) between
the different black hole metrics.
\\

{\bf Acknowledgements :}  We thank the referee for constructive comments that helped to improve the manuscript.

\end{document}